\documentclass[epj,nopacs]{svjour}
\usepackage[british]{babel}
\usepackage[intlimits,tbtags]{amsmath}
\usepackage{graphics}
\usepackage{epsfig}
\usepackage{appendix}
\usepackage{color}

\begin{document}
\hugehead
\title{Experimental access to Transition Distribution Amplitudes
with the \={P}ANDA experiment at FAIR}
\author{{\large The \={P}ANDA Collaboration} \\ \\
B.P.~Singh \inst{1}
\and W.~Erni \inst{2}
\and I.~Keshelashvili \inst{2}
\and B.~Krusche \inst{2}
\and M.~Steinacher \inst{2}
\and B.~Liu \inst{3}
\and H.~Liu \inst{3}
\and Z.~Liu \inst{3}
\and X.~Shen \inst{3}
\and C.~Wang \inst{3}
% ...
}
\institute{Aligarth Muslim University, Physics Department, Aligarth India
\and Universit\"at Basel  Switzerland
\and Institute of High Energy Physics, Chinese Academy of Sciences, Beijing  China
% ...
}
\date{Received: date / Revised version: date}
%
%
\abstract{
Baryon-to-meson Transition Distribution Amplitudes (TDAs) encoding valuable new information
on hadron structure appear as building blocks in the collinear factorized description for
several types of hard exclusive reactions.
In this paper, we address the possibility of accessing
nucleon-to-pion ($\pi N$) TDAs from $\bar{p}p \to e^+e^- \pi^0$ reaction
with the future \={P}ANDA detector at the FAIR facility.
At high center of mass energy and high invariant mass squared of the lepton pair $q^2$,
the amplitude of the signal channel $\bar{p}p \to e^+e^- \pi^0$
admits a QCD factorized description
in terms of $\pi N$ TDAs and nucleon Distribution Amplitudes (DAs) in the forward and backward
kinematic regimes. Assuming the validity of this factorized description,
we perform  feasibility studies for measuring $\bar{p}p \to e^+e^- \pi^0$
with the \={P}ANDA detector.
Detailed simulations on signal reconstruction efficiency as well as on rejection of the most
severe background channel, {\it i.e.} $\bar{p}p \to \pi^+\pi^- \pi^0$
were performed for the center of mass energy squared $s = 5$ GeV$^2$ and $s = 10$ GeV$^2$,
in the kinematic regions $3.0 < q^2 < 4.3$ GeV$^2$ and $5 < q^2 < 9$ GeV$^2$,
respectively, with a neutral pion scattered in the forward or backward cone
$| \cos\theta_{\pi^0}| > 0.5 $ in the proton-antiproton center of mass frame.
Results of the simulation show that the particle identification capabilities
of the \={P}ANDA detector will allow to achieve a background rejection factor
of $5\cdot 10^7$ ($1\cdot 10^7$) at low (high) $q^2$ for $s=5$ GeV$^2$, and
of $1\cdot 10^8$ ($6\cdot 10^6$) at low (high) $q^2$ for $s=10$ GeV$^2$,
while keeping the signal reconstruction efficiency at around $40\%$.
At both energies, a clean lepton signal can be reconstructed with the expected statistics
corresponding to $2$ fb$^{-1}$ of integrated luminosity.
The cross sections obtained from the simulations are used
to show that a test of QCD collinear factorization can be done at the lowest order
by measuring scaling laws and angular distributions.
The future measurement of the signal channel cross section with \={P}ANDA
will provide a new test of the perturbative QCD description of a novel class of
hard exclusive reactions and will open the possibility
of experimentally accessing $\pi N$ TDAs.
%
} 
%
%
\maketitle
\section{Introduction}
\label{sec:intro}
Studies of hard exclusive reactions, such as Deeply Virtual Compton Scattering (DVCS) 
and Hard Exclusive Meson Electroproduction, within the collinear factorization approach, 
allow to challenge a QCD-based description of hadron structure 
(for a review see {\it e.g.} \cite{Boffi:2007yc}).
By separating the hard and soft stages of the interaction,
at high energies the amplitudes of these reactions can be presented in the form
of convolutions of hard parts, computable in perturbation theory, and soft parts:
generalized parton distributions (GPDs) and meson distribution amplitudes (DAs).
These non-perturbative objects can be assigned a rigorous meaning in QCD and allow
to interpret hadronic structural information in terms of quark and gluon degrees of freedom.
Along with the usual parton distribution functions (PDFs) and form factors (FFs),
GPDs encode valuable structural information about hadrons.
In particular, GPDs are currently seen
as a tool to study the nature and origin of the nucleon spin. Moreover, GPDs allow an
extremely vivid interpretation in the impact parameter space as spatial femto-photographs
of the hadron interior in the transverse plane.

Further development of the GPD approach led to the introduction of baryon-to-meson
transition distribution amplitudes (TDAs) \cite{Strikman,Pire:2004ie}
broadening the class of hard reactions for which
a factorized description of the  scattering amplitudes for strong interaction phenomena
can be applied. The physical picture encoded in baryon-to-meson TDAs is conceptually close
to that contained in baryon GPDs and baryon DAs. Baryon-to-meson TDAs probe partonic
correlations between states of different baryonic charge thus giving access to non-minimal
Fock components of baryon light-cone wave functions. Fourier transforming TDAs
to the impact parameter space allows one to perform femto-photography
of hadrons from a new perspective. In particular, nucleon-to-pion ($\pi N$) TDAs may be used
as a tool for spatial imaging of the structure of the pion cloud inside the nucleon.
This opens a new window for the investigation of the various facets
of the nucleon internal structure.
A dedicated program for accessing $\pi N$ TDAs in the space-like regime
through backward pion electroproduction \cite{Lansberg:2007ec,Lansberg:2011aa}
was proposed for JLab Hall B $@$ $12$ GeV
(see Ref.~\cite{KubaAlex} for preliminary studies
dedicated to JLab $@$ $6$ GeV).

The future \={P}ANDA (antiProton ANnihilations at DArmstadt) experiment at FAIR
(Facility for Antiproton and Ion Research)
operating a high-intensity antiproton beam with momentum up to $15$ GeV
offers unique possibilities for new investigations of the hadron structure
(see Refs. \cite{PANDAprogramm,Wiedner:2011mf}) complementing the results obtained
from the studies of lepton beam induced reactions. In particular, the \={P}ANDA
experimental program includes dedicated measurements of the time-like
electromagnetic form factors of the proton, mainly through the annihilation process
$\bar{p} p \to e^+ e^-$, for which feasibility studies with the \={P}ANDA  detector
have already been performed  at several antiproton beam energies \cite{Sudol:2009vc}.
The high intensity of the antiproton beam,
together with the performance of the \={P}ANDA  detector,
including particle identification capabilities, will render an unprecedent accuracy
in the measurements over a large range of four-momentum transfer squared,
as shown by the simulations.

Outside the resonance region
({\it i.e.} for sufficiently high invariant mass of the lepton pair)
the nucleon electromagnetic form factors admit a factorized description
within the perturbative QCD (pQCD) approach and follow a scaling law
\cite{Lepage:1980,Chernyak:1983ej}. This framework was further developed in
\cite{Pire:2004ie,Pire:2005ax} and was employed in 
\cite{Lansberg:2007se,Lansberg:2012ha,Pire:2013tpa}
to provide a factorized description of nucleon-an\-ti\-nu\-cleon annihilation
into a highly virtual lepton pair and a pion in terms of $\pi N$ TDAs and nucleon DAs.
Note that a similar treatment can be applied to the scattering amplitude, when the lepton pair
originates from a heavy charmonium state \cite{Pire:2013jva,Ma:2014pka,Ma:thesis}.
At lower energies, where factorization does not hold, descriptions
of the $\bar{p}p \to e^+ e^- \pi^0$ amplitude
in terms of a one-nucleon-exchange model and the Regge theory
\cite{Adamuscin:2006bk,Guttmann:2012sq} have been proposed,
and preliminary studies of the cross section measurement with \={P}ANDA
have already been performed \cite{Boucher}.

Thus, alongside with the time-like electromagnetic 
form factor measurements, it is extremely appealing to test the predictions 
of the pQCD collinear factorized description of $\bar{p}p \to e^+ e^- \pi^0$ 
and address the possibility of accessing the proton/antiproton-to-pion TDAs 
with the \={P}ANDA detector through the measurement of the corresponding 
differential cross section.

In the present paper we consider the feasibility of measuring
the $\bar{p}p \to e^+ e^- \pi^0$ signal channel cross sections at high center-of-mass energy 
and high four-momentum squared of the virtual photon, with the produced $\pi^0$ scattered 
into the forward or backward angular regions, in which the factorization theorem 
is expected to be valid \cite{Mamen_thesis}.
Proton-antiproton annihilation into three pions,
{\it i.e.} $\bar{p}p \to \pi^+ \pi^- \pi^0$,
appears as the most severe background channel for the
process of interest, as it contains the same number of particles in the
final state, with identical charge signature.
Detailed simulations have been performed on
the signal reconstruction efficiency and on the background rejection.
The feasibility of the measurement has been studied
using an integrated luminosity of $2$~fb$^{-1}$.
The cross sections obtained from the simulations are used
to test pQCD at the lowest order by measuring scaling laws 
and angular distributions.

\section{Set-up of the future \={P}ANDA experiment}
\label{sec:detector}
An extensive description of the \={P}ANDA detector
can be found in Ref.~\cite{PANDAprogramm}.
Here we give a brief outline of the main components
which are relevant to this analysis.

A High Energy Storage Ring (HESR),
in which both stochastic and electron cooling systems are foreseen,
will provide a high quality antiproton beam of momentum between 1.5 and 15 GeV.
The concept of the detector, the read out and the data acquisition system
are similar to that of other recently built detectors,
such as ATLAS, CMS, COMPASS and BaBar.
However, the high expected rate of 2$\cdot 10^7$ interactions per second
and the multipurpose character of the detector,
including the measurement of low cross sections in the charm sector,
demand unique detection capabilities in \={P}ANDA.
These include geometrical acceptance of almost $4\pi$,
energy and momentum resolutions at the level of a few percent,
fast data acquisition and high radiation hardness.
In the HESR high-luminosity mode,
the average design luminosity of ${\cal L}= 1.5 \cdot 10^{32}$ cm$^{-2}\hspace{1.0pt}$s$^{-1}$
will be reached with a pellet target of thickness $4\cdot 10^{15}$ hydrogen atoms/cm$^{2}$,
and $10^{11}$ stored antiprotons in HESR.
The detector is divided in a target spectrometer,
in which the target is surrounded by a solenoid magnet providing
up to 2\,T magnetic field, and a forward spectrometer, based on a 2\,Tm
dipole magnet, to ensure particle detection at small polar angles,
down to $2^{\circ}$.
Tracking, particle identification, electromagnetic calorimetry,
and muon identification detectors are designed for both spectrometers.
The reconstruction of the interaction point
as well as secondary vertices is done with the microvertex detector (MVD).
The concept of the MVD is based on radiation hard silicon pixel detectors
with fast individual pixel readout circuits and silicon strip detectors,
making a four layer barrel detector with an inner radius of $2.5$ cm
and an outer radius of $13$ cm.
The charged particle tracking and identification is provided
by the straw tube tracker (STT), consisting of
aluminized mylar tubes called ``straws'',
arranged in planar layers and mounted around the MVD
in a total of $24$ layers. Of these, the $8$ central ones
are tilted to achieve a resolution of $3$ mm also in the direction
parallel to the beam.
Track detection at angles below $22^{\circ}$ (not fully covered by the STT)
is completed by three chambers of gas electron multiplier (GEM) detectors
placed $1.1$ m, $1.4$ m and  $1.9$ m downstream of the target.
The chambers are designed to sustain a high counting rate
of particles peaked at the most forward angles
due to the relativistic boost of the reaction products.
Additional components are required for the identification of
hadrons and leptons in a wide kinematic range.
For slow particles at large polar angles,
particle identification will be provided
by the time-of-flight (TOF) detector,
with time resolution between $50$ and $100$ ps
as required by the $50-100$ cm of flight path
in the target spectrometer.
The PbWO$_4$ electromagnetic calorimeter (EMC), operated at $-25^{\circ}$ C,
is designed for the detection of photons and electrons.
Here, the fast scintillator material with short radiation length
is required by the expected high counting rates and
geometrically compact design of the target spectrometer.
Crystals of $20$ cm length, i.e. approximately 22 radiation lengths,
are used in order to achieve an energy resolution
below $2\%$ at $1~\rm GeV$.
For the efficient separation of pions from electrons at momenta $p<1$\,GeV,
a barrel and a forward disk DIRC
(detection of internally reflected Cherenkov light)
complete the particle identification (PID) system.

\section{PANDA detector reconstruction capabilities}
\label{sec:detector_simulation}
In the physics analysis, the generated events by the Monte Carlo programs
(corresponding to signal or background channels) are, in a first step,
passed through a full simulation
of the \={P}ANDA  detector, based on the {\sc Geant 4} package~\cite{Agostinelli:2002hh},
which takes care of the propagation of particles through the detector.
Hit and energy loss information is then digitized
according to a model simulating electronic properties,
including electronic noise, yielding a response of the different detectors.
The second step is the reconstruction of the relevant physical quantities
for the identification of electrons, such as momentum, ratio of energy loss
to path length $dE/dx$ in the STT,
Cherenkov angle in the DIRC detectors,
and energy deposit in the EMC
from the simulated data.
These two steps have been described in detail in Ref.~\cite{PANDAprogramm},
so we will give here only the main features which are relevant
for the electron and photon identification.

The truncated arithmetic mean method is used on the $dE/dx$ values
for particle identification in order to exclude from
the sample the largest values which correspond to the extended Landau tail
of the distribution. The value used for the calculation of the arithmetic mean
corresponds to 70\% of the $N$ individual $dE/dx$ values.
In this way a compromise between the requirements of the best resolution,
defined through the width of a Gaussian fit, and the smallest tail
of the distribution is achieved. A resolution of $<10\%$ in $dE/dx$
is obtained for pions of momentum 1\,GeV,
which corresponds on average to four standard deviations
of the distance between the truncated means for electrons and pions.

For the DIRC detector, the Cherenkov angle is given with
a resolution $ \sigma_{C}=\sigma_{C,\gamma}/\sqrt {N_{ph}}$,
where the single photon resolution is $\sigma_{C,\gamma}=10$\,mrad.
The number of detected photons, $N_{ph}$, has a dependence on the velocity
and path length of the particle travelling inside the Cherenkov radiator.
To calculate the Cherenkov angle, the software
takes also into account the quantum efficiency of the photodectectors
and the transmission and reflectivity losses in the detector material.
A resolution of 2.3\,mrad is obtained for
pions of momentum 1\,GeV \cite{Fohl:2008zza}. The DIRC discrimination power
is higher at lower energies due to the larger difference between the Cherenkov angles
for pions and electrons: at momentum 500 MeV the difference in the angles
for the Cherenkov light amounts to 36\,mrad whereas at momentum 1.5 GeV
it is 4\,mrad.

\begin{figure*}[hbtp]
\begin{center}
\epsfig{file=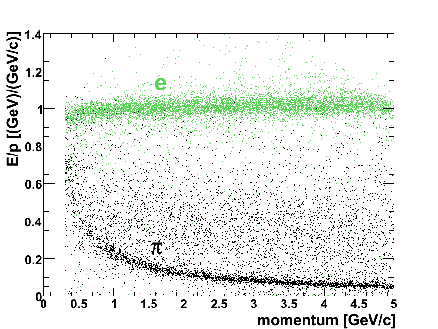,width=8.5cm}
\epsfig{file=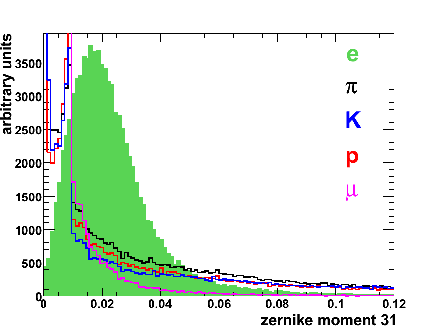,width=8.5cm}
\caption
[\protect{}]
{
The ratio $E/p$ between the measured energy deposit $E$
and the reconstructed momentum $p$ for an electron sample
and for a pion sample (left) and the distribution of the Zernike moment $31$
for samples of different particle species (right; colour is available 
in the electronic version of the paper).
The figures are taken from Refs.~\cite{PANDAprogramm,Sudol:2009vc}.
}
\label{fig:pid}
\end{center}
\end{figure*}

The most important detector for electron identification is
the electromagnetic calorimeter. Electron identification is done
using the ratio $E/p$ between the measured energy deposit $E$
and the reconstructed momentum $p$.
In the electromagnetic calorimeter, the electrons
deposit all (up to minor losses due to dead material, crystal edges, etc.)
their energy via an electromagnetic shower, whereas
muons and hadrons loose only a much lower fraction of their energy
via Bethe-Bloch excitations and ionization processes.
However, there could be cases in which
a high energy deposit would be the consequence of hadronic interactions.
In those cases the analysis of the shower shape plays an important role
in the particle identification process.
The Moli\`ere radius of PbWO$_4$ is 2\,cm and it is of the order
of the crystal front size dimensions, $2.1\times2.1$\,cm$^2$ in the barrel and backward endcap 
and $2.44 \times 2.44$\,cm$^2$ in the forward endcap of the calorimeter.
In the case of an electromagnetic shower the largest fraction
of the energy deposition is contained in a few crystals,
whereas in the case of a hadronic interaction, the energy deposition
will be distributed in a larger volume. The shower shape analysis uses
the energy deposited in the central crystal of the cluster
relative to that in the $3\times 3$ or $5\times 5$ crystal arrays
surrounding it. The ratio between these two numbers
is a measure for the cluster size and shape, and therefore it is an indicator
for an electromagnetic or a hadronic interaction.
In addition, a set of four Zernike moments\footnote{The Zernike polynomials
are a complete orthogonal set in the unit disk $0 < x^2 + y^2 < 1$.
The projections of a function $f(x,y)$ on the basis of the Zernike polynomials
are called the Zernike moments of $f$. Details can be found, for instance,
in Ref.~\cite{Zernike}, chapter $9$, section $2$.}
are used to describe the spatial distribution of the energy
within the shower by using polynomials
in the radial and angular coordinates.
Fig.~\ref{fig:pid} shows two examples on how $E/p$ and
one of the Zernike moments can be used to discriminate
electrons from pions (for an extensive description,
see Ref.~\cite{PANDAprogramm}, chapter 3, subsection 3.3.3).

Probabilities for the identification of a given particle
using different hypotheses (electron, muon, pion, kaon and proton)
are calculated on the basis of the results given by simulations using
these species as input for the event generators for an extended
range of momenta and polar angles.
In addition to the variables discussed above,
$dE/dx$ information from the microvertex detector
as well as hit information from the muon detector are included.
In the case of the electromagnetic calorimeter,
this probability is calculated using the output of a neural network
which uses as the input the list of shower shape and Zernike parameters
for a cluster described previously, as discussed in~\cite{PANDAprogramm}.
A global particle identification
likelihood can be calculated using the individual subdetector
likelihoods. Depending on the signal and background channels,
the cuts for the particle identification can be adjusted
to get the best signal efficiency for the required background suppression.

In this analysis we use a number of simplifications with respect to
the continuously developing \={P}ANDA framework.
Charged particle tracking was performed without pattern recognition, leading to an
overestimation of the track finding efficiency compared to the performance
studies summarized in Ref~\cite{Erni:2013ita}. 
The Kalman filter for track fitting used a less refined material distribution. 
For the Cherenkov angle, photon transport and photon detection was not simulated, 
but instead a smearing technique was applied. 
For the description of the electromagnetic showers {\sc Geant 4.7} was used,
for which deviations to data were reported by the BaBar experiment~\cite{Banerjee:2008zzb}.
\begin{figure*}[hbtp]
\begin{center}
\epsfig{file=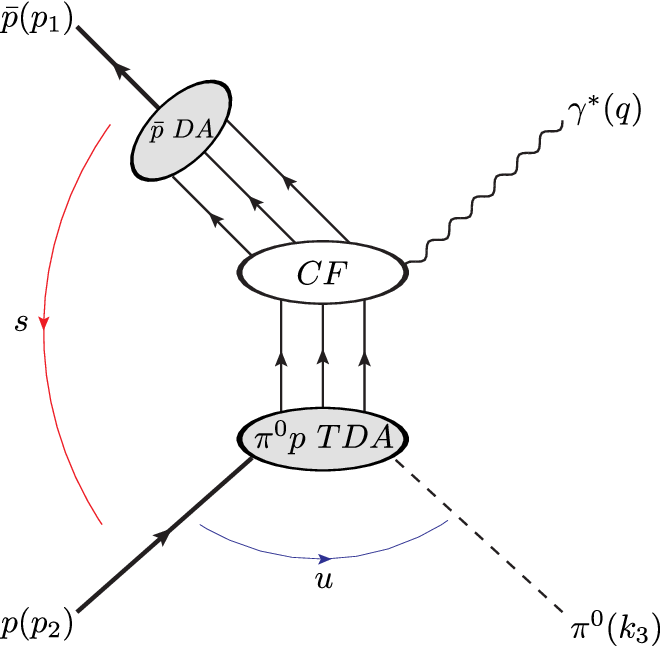,width=7.5cm}
\hspace*{1.5cm}
\epsfig{file=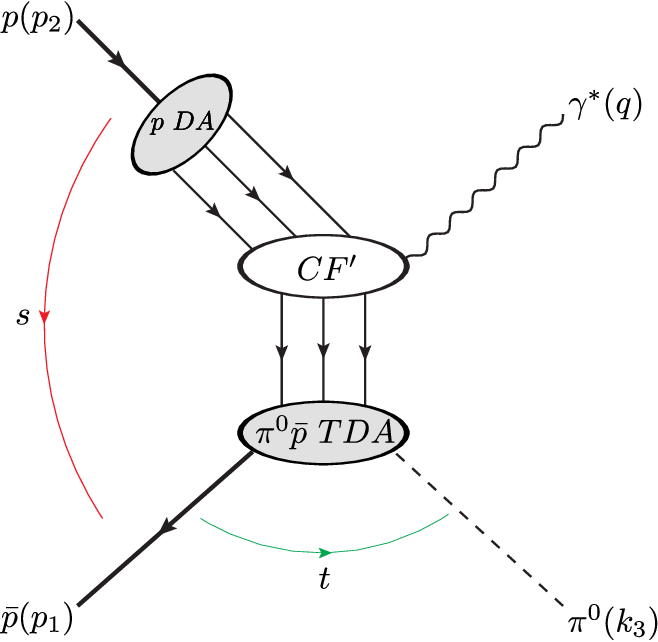,width=7.5cm}
\caption
[\protect{}]
{
The two possibilities for factorization in the annihilation process
$\bar{p}p \rightarrow \gamma^* \pi_0$, for both kinematics: backwards (left)
and forward (right). $\bar{p}$ ($p$) DA stands for
the distribution amplitude of antiproton (proton).
$\pi^0 p$ ($\pi^0 \bar{p}$) TDA stands for
the transition distribution amplitude from a proton (antiproton)
to a neutral pion. $CF$ and $CF^{\prime}$ stand for coefficient functions
(hard subprocess amplitudes).
}
\label{fig:diagrams}
\end{center}
\end{figure*}

\section{Theoretical overview and event generation}
\label{sec:theory}
In this section we present a short overview of the basic definitions and conventions
employed for the factorized description of the nucleon-antinucleon annihilation
into a high invariant mass lepton pair in association with a
$\pi^0$ meson. The details can be found in
Refs.~\cite{Lansberg:2007se,Lansberg:2012ha,Pire:2013tpa}.

To the leading order in the electromagnetic coupling the reaction proceeds in two stages:
firstly proton and antiproton annihilate to produce a virtual photon and a neutral pion
and subsequently the virtual photon decays into the lepton pair:
\begin{equation}
\begin{split}
\bar{p} (p_1,s_1)+  p (p_2,s_2) &\rightarrow \gamma^*(q)+ {\pi^0}(k_{3}) \\
&\rightarrow e^+(k_1) + e^-(k_2) +{\pi^0}(k_{3}) ~,
\label{BarNNannihilation reaction}
\end{split}
\end{equation}
where by $s_{1,2}$ we denote the antinucleon and nucleon spin variables.

According to the usual \={P}ANDA conventions, we choose the $z$
axis along the colliding $\bar{p}p$ with the positive direction along the antinucleon beam.
The two remaining  spatial directions are referred to as the transverse plane.
In order to specify the two kinematic regimes subject to the factorized description
in terms of $\pi N$ TDAs we switch to the light-cone variables and introduce the
$t$- and $u$-channel light-cone vectors
$n^t$, $p^t$;  $n^u$, $p^u$ ($p^2=n^2=0$, $2 p \cdot n=1$).
To quantify the longitudinal momentum transfers in the appropriate channels
we define the $t$- and $u$-channel skewness variables
\begin{equation}
 \xi^t \equiv - \frac{(k_{3}-p_1) \cdot n^t}{(k_{3}+p_1) \cdot n^t} \qquad
 \xi^u \equiv - \frac{(k_{3}-p_2) \cdot n^u}{(k_{3}+p_2) \cdot n^u} ~~.
\end{equation}

The factorization mechanism suggested in \cite{Lansberg:2007se}
for the $\bar{p} (p_1)+  p (p_2) \rightarrow \gamma^*(q)+ {\pi^0}(k_{3})$
subprocess of the reaction (\ref{BarNNannihilation reaction})
is schematically depicted on Fig.~\ref{fig:diagrams}. 
The amplitude is presented as a convolution
of the hard part computed by means of perturbative QCD with nucleon DAs and
nucleon-to-pion TDAs encoding the soft dynamics. The factorization is assumed to be achieved
in two distinct kinematic regimes:
\begin{itemize}
\item  the near forward regime
($s=(  p_1+p_2)^2 $, $q^2$ - large with $\xi^t$ fixed; and $|t|=|(k_3-p_{1})^2| \sim 0$);
it corresponds to the produced pion moving nearly in the direction of the initial
$\bar{p}$ in the $\bar{p}p$ center-of-mass (CM) system.
\item  the near backward regime
($s=(  p_1+p_2)^2 $, $ q^2$ - large with $\xi^u$ fixed; and $|u|=|(k_3-p_2)^2| \sim 0$);
it corresponds to the produced pion moving nearly in the direction of the initial
$p$ in $\bar{p}p$ CM system.
\end{itemize}
The suggested reaction mechanism should manifest itself through the distinctive forward
and backward peaks of the $\bar{p} p\to \gamma^* \pi^0$ cross section.
The charge conjugation invariance results in the perfect symmetry between 
the two kinematic regimes.
In what follows, for definiteness, we focus on the near forward kinematic regime.
From now on we omit the labels referring to the particular ($t$- or $u$-) kinematic regime.
However, all formulas for the near backward kinematics are essentially the same as
in the forward kinematics (after interchanging the momenta).
To  the leading twist accuracy  and to the leading order in
the strong coupling $\alpha_s$, the amplitude $\mathcal{M}_{\lambda}^{s_1 s_2}$
of $\bar{p} p \to \gamma^* \pi^0$ reads
\begin{equation}
\mathcal{M}_{\lambda}^{s_1 s_2}=
\mathcal{C} \frac{1}{(q^2)^2}
\Big[
\mathcal{S}_{\lambda}^{s_1 s_2}
\mathcal{I}(\xi, t)-
\mathcal{S'}_{\lambda}^{s_1 s_2}
\mathcal{I}'(\xi, t)
\Big] ~,
\label{Def_ampl_M}
\end{equation}
where
\begin{eqnarray}
{ \cal C}=-i \frac{(4 \pi \alpha_s)^2 \sqrt{4 \pi \alpha_{em}} f_N^2}{54 f_\pi}~.
\end{eqnarray}
Here $\alpha_s$ and $\alpha_{em}$ are the strong and electromagnetic coupling constants,
$f_N$ stands for the nucleon wave function normalization constant,
and $f_\pi=93$ MeV denotes the pion weak decay constant.
The spin structures in Eq.~(\ref{Def_ampl_M}) are defined as
\begin{eqnarray}
&&
\mathcal{S}_{\lambda}^{s_1 s_{2}}  \equiv
\bar{V}(p_1,s_{1}) \hat{\varepsilon}^*(\lambda) \gamma_5 U(p_2,s_2)
\nonumber \\ &&
\mathcal{S'}_{\lambda}^{s_1 s_{2}} \equiv
\frac{1}{M}
\bar{V}(p_{1},s_{1}) \hat{\varepsilon}^*(\lambda) \hat{\Delta}_T \gamma_5 U(p_2,s_2) ~,
\end{eqnarray}
where $V$ and $U$ are the usual nucleon Dirac spinors;
$\Delta_T \equiv (k_3-p_1)_T$ denotes the transverse $t$-channel momentum transfer
and the Dirac ``hat'' notation $\hat{v}= \gamma_\mu v^\mu$ is employed.
$\varepsilon(\lambda)$ stands for the polarization vector of the virtual photon.
$\mathcal{I}$ and $\mathcal{I'}$ denote the convolution integrals of $\pi N$
TDAs and nucleon DAs with the hard scattering kernels computed from the set of
relevant scattering diagrams \cite{Lansberg:2007ec}.
The averaged-squared amplitude for the process (\ref{BarNNannihilation reaction}) then reads
\begin{equation}
\begin{split}
&|\overline{\mathcal{M}^{   \bar{p}p \rightarrow e^+e^- \pi^0 }}|^2 = \\
&\frac{1}{4}
\sum_{s_1, \, s_{2}, \, \lambda, \, \lambda'}
\mathcal{M}_{\lambda}^{s_1 s_{2}}
\frac{1}{q^2}
e^2 {\rm Tr}
\left\{
\hat{k}_2
 \hat{\varepsilon}(\lambda)
 \hat{k}_1
 \hat{\varepsilon}^*(\lambda')
\right\}
\frac{1}{q^2}
\left( \mathcal{M}_{\lambda'}^{s_1 s_{2}} \right)^* ~.
\label{WF_Cross_sec}
\end{split}
\end{equation}
The differential cross section of the reaction
(\ref{BarNNannihilation reaction}) is expressed as
\begin{equation}
\frac{d \sigma}{dt ~dq^2 ~d\cos\theta_{\ell}^*}=
 \frac{\int d \varphi_\ell^* \, |\overline{\mathcal{M}^{ \bar{p} p\rightarrow e^+ e^- \pi^0}}|^2 }
 {64 s (s-4M^2) (2 \pi)^4} ~,
 \label{Cross_sec_PANDA}
\end{equation}
where $\theta_{\ell}^*$ and $\varphi_\ell^*$
are the lepton polar and azimuthal angles defined in
the $e^+ e^-$ CM frame ({\it i.e.} the $\gamma^*$ rest frame).

To the leading twist accuracy, only the transverse polarization states of the virtual photon
are contributing. Computing the relevant traces
and integrating over the lepton azimuthal angle one gets
\begin{equation}
\begin{split}
 \int d \varphi_{\ell}^* \, &|\overline{\mathcal{M}^{ \bar{p} p \rightarrow e^+ e^- \pi^0 }}|^2
\Big|_{\rm Leading \, twist} \\
&= 2 \pi e^2(1+\cos^2 \theta_\ell^*)  \frac{1}{4} |\mathcal{C}|^2 \frac{2(1+\xi)}{\xi (q^2)^4} \\
& ~~\times \big( |\mathcal{I}(\xi, t)|^2- \frac{\Delta_T^2}{M^2} |\mathcal{I}'(\xi, t)|^2 \big)~.
\end{split}
\end{equation}
Neglecting $t$ and the nucleon mass squared $M^2$ with respect to large invariants $s$ and $q^2$
(that is a reasonable approximation in the kinematic domain in which the
factorized description is assumed to hold) the skewness parameter can be expressed as
\begin{equation}
\xi \simeq \frac{q^2}{2 s -q^2} ~.
\end{equation}

Thus, we work out the following  expression for the differential cross section of the reaction
(\ref{BarNNannihilation reaction}) within the factorized description in terms of $\pi N$ TDAs
in the near-forward kinematic regime:
\begin{equation}
\begin{split}
&\frac{d \sigma}{dt ~d q^2 ~d \cos \theta_{\ell}^*} \Big|_{\rm Leading \, twist} =  \\
& \qquad\qquad \frac{K}{s-4M^2} \frac{1}{(q^2)^5} (1+\cos^2 \theta_\ell^*) ~,
\label{CS_working fromula}
\end{split}
\end{equation}
where
\begin{equation}
\begin{split}
K=&\frac{(4 \pi \alpha_{em})^2 (4 \pi \alpha_{s})^4 f_N^4}{64 \cdot  54^2 (2 \pi)^3 f_\pi^2} \\
& \times \big( |\mathcal{I}(\xi, t)|^2- \frac{\Delta_T^2}{M^2} |\mathcal{I}'(\xi, t)|^2 \big) ~.
\end{split}
\end{equation}

To compute the integral convolution $\cal I$, ${\cal I}'$ we use the
revised version of the phenomenological model for $\pi N$ TDAs suggested in
Refs. \cite{Lansberg:2007se,Pire:2011xv}.
Within this approach $\pi N$ TDAs are constrained from the chiral dynamics
and expressed through the nucleon DAs relying on the soft pion theorem.
Certainly, this is an oversimplified $\pi N$ TDA model that gives non-zero contribution
only into the convolution $\mathcal{I}$. Moreover, within this model $\mathcal{I}$
turns to be $\xi$- and $t$- independent.
Nevertheless, this model is supposed to provide a reasonable estimate
of the normalization for $\pi N$ TDAs and can be taken as reliable at least
for sufficiently small transverse momentum transfer.
We refer the reader to Ref.~\cite{Stefanis:1999wy} for the discussion on various 
phenomenological solutions for the nucleon DA and the relevant values of the strong coupling 
and nucleon wave function normalization constant. In the present analysis we use 
the Chernyak-Ogloblin-Zhitnitsky (COZ) \cite{Chernyak:1987nv}
phenomenological solution for the nucleon DAs. This solution yields the value
$|\mathcal{I}|^2=1.69 \cdot 10^9$, which is used in our evaluation.
For the numerical estimates we use, following Ref.~\cite{Lansberg:2012ha},
the mean value of the strong coupling $\alpha_s=0.3$ and $f_N = 5.2 \cdot 10^{-3}$ GeV$^2$.
The cross section (\ref{CS_working fromula}) serves as the input for the event generator
of the signal events $\bar{p}p \to e^+ e^- \pi^0$ whose source code \cite{Pelizaeus}
was interfaced to the EvtGen \cite{Ryd} Monte Carlo.
We remark that the cross section (\ref{CS_working fromula}) does not
contain QED radiative corrections, so the PHOTOS package~\cite{Barberio:1993qi}
has been  consistently switched off in the {\sc Geant} simulation.

For the cross section of the most severe background channel,
i.e. three-pion production $\bar{p} p \to \pi^+ \pi^- \pi^0$,
no theoretical calculations in the kinematic region of interest are available
and the few existing low-precision measurements
\cite{Bacon:1973ax,Czyzewski:1963,Everett:1974ni,Sai:1982dv,Abele:1999ac}
are not sufficient to constrain models.
Inspired by the expectation for the total cross section ratio
$\sigma(\bar{p} p \to \pi^+ \pi^-)/ \sigma(\bar{p} p \to e^+ e^-) \sim 10^6$
(see \cite{Sudol:2009vc,VandeWiele:2010kz} and references therein),
we have assumed that the same relation holds for the case
$\sigma(\bar{p} p \to \pi^+ \pi^- \pi^0)/ \sigma(\bar{p} p \to e^+ e^- \pi^0)$.
Even when data sets suggest that three-pion production is about an order of magnitude
higher than two-pion production, the totally unknown $\bar{p} p \to e^+ e^- \pi^0$
cross section supports the assumption on the signal to background ratio.
We remark that in this analysis we reach a very small background pollution
on the signal sample. The precise value of the ratio
$\sigma(\bar{p} p \to \pi^+ \pi^-)/ \sigma(\bar{p} p \to e^+ e^-)$ 
is not critical for the conclusions.
When the \={P}ANDA experiment is running,
the measurement of the $\pi^+\pi^-\pi^0$ cross section will be done with great precision,
and simultaneously to that of the $e^+e^-\pi^0$ events, so the three-pion cross section
will be available to perform background subtraction precisely.
In addition, we have assumed that the angular distributions for the three-pion final state
$\pi^+ \pi^- \pi^0$ are identical to that of the signal final state
$e^+ e^- \pi^0$. With these considerations in mind, in the event generator for signal events,
lepton masses and Monte Carlo identifiers were replaced by the ones correponding to pions
to account for background production.
This conservative approach represents, from the experimental point of
view, the most unfavored situation for background rejection. Having identical distributions
for signal and background then requires to rely entirely on particle identification for the
discrimination of signal and background events.

Only $\bar{p}p \rightarrow \pi^+ \pi^- \pi^0$ background events have been
simulated in this analysis. As already stated, three-pion production,
having the same number of final state particles as that of the signal channel,
the same charge signature and together with the small mass gap between electron an pion, 
constitutes the most severe channel in terms of suppression. 
Moreover, the assumption of identical angular distributions
for signal and background events becomes the worst possible scenario 
as rejection concerns. In the spirit of a first feasibility study, 
other possible sources of background are left for future investigations,
since their contribution to the signal pollution is estimated to be minor in comparison
with three-pion production. Simple cross section estimations have been done
with the help of the Dual Parton Model (DPM) Monte Carlo~\cite{Galoian:2005ea}
for some additional background channels. 
For a compilation of the existing data sets on $\bar p$ reactions, see, 
for instance, Ref.~\cite{Flaminio-84}.
The production cross section for $\bar{p}p \rightarrow K^+K^-\pi^0$ is roughly 
two orders of magnitude smaller than the cross section for the simulated channel
$\bar{p}p \rightarrow \pi^+\pi^-\pi^0$ over an extended range of $\bar{p}p$ 
center of mass energies. In addition, kaons are much better separable from electrons 
than pions due to the larger mass gap by means of kinematical fits.  
The cross section for $\bar{p} p \rightarrow \pi^0 \pi^0$
is roughly $30$ times smaller than that of $\bar{p} p \rightarrow \pi^+\pi^- \pi^0$.
In $\bar{p} p \rightarrow \pi^0 \pi^0$, one of the two pions can undergo
a Dalitz decay $\pi^0 \rightarrow e^+e^-\gamma$, which could fake a signal signature.
The Dalitz decay has a branching ratio of $1.2\%$~\cite{PDG2014},
and thus is suppressed by a factor $2500$ compared to $\pi^+\pi^- \pi^0$.
Moreover, it has an additional photon.  
These background events can not be separated from signal events
by means of PID only, and additional kinematic cuts will have to be developed
if further suppression is needed. 
Studies of the non-resonant background are beyond the scope of the present analysis.
These include exclusive QED channels, like $\bar{p} p \rightarrow e^+e^-\gamma\gamma$,
as well as photons from uncorrelated events in coincidence with 
$\bar{p} p \rightarrow e^+ e^-$ within the data acquisition window.
For the latter, dedicated full simulations are needed to make reliable estimates.

One of the key problems we need to address is the experimental 
verification of the validity of the pQCD-factorization assumption
for the reaction in question. Providing evidence of the applicability 
of the factorized description at relatively low values of $q^2$ 
represents the most important potential physical result of the
suggested measurements. From the theory side (see {\it e.g.} Ref.~\cite{Collins:1996fb}) 
several essential marking signs exist for the onset
of the collinear factorization regime for hard exclusive reactions:
\begin{itemize}
\item Dominance of the specific polarization of the virtual photon.
\item Characteristic scaling behaviour of the cross section in $1/q^2$.
\item Universality of the corresponding non-perturbative quantities, which means 
that the same non-perturbative objects provide a satisfactory description to several
hard exclusive reactions. 
\end{itemize}

For the case of the nucleon-antinucleon annihilation into
a lepton pair in association with a forward (or backward) neutral
pion it is the transverse polarization of the virtual photon that is dominant
within the collinear factorized description in terms of $\pi N$ TDAs.
This dominating contribution manifests itself through the characteristic 
$(1+\cos^2 \theta_\ell^*)$ behaviour of the cross section 
({\it c.f.} Eq.~(\ref{CS_working fromula})).
This term can be extracted from the experimentally measured cross section through 
the harmonic analysis in the lepton pair CM scattering angle $\theta_\ell^*$.
The characteristic $q^2$-scaling behaviour of the cross section is explicit 
from Eq.~(\ref{CS_working fromula}).

It is worth mentioning that probing experimentally the validity
of the collinear factorization assumption for hard exclusive reactions usually 
represents a challenging task. 
The test of the scaling behaviour with the available 
small lever arm in $q^2$ (that is typical for the fixed-target kinematics experiments)
turns out to be intricate due to the uncontrollable higher-twist contributions 
and model dependent implementation of skewness dependence within a particular model 
for the relevant non-perturbative objects (GPDs or TDAs).
Testing factorization will then demand the use of NLO QCD fits to
the $q^2$-dependence of the cross section to separate the contributions 
of the longitudinal and transverse photon polarizations.
In a similar way,  detailed harmonic analysis will be needed to discriminate 
between different Fourier components of the $\cos\theta^*_{\ell}$ distributions.
To illustrate these difficulties consider the controversial issue 
of the applicability of the GPD-formalism-based description of 
near-forward hard exclusive pion electroproduction off protons.
Existing data are suggestive, but not conclusive and the consistency 
of factorized description still remains to be shown within the 
experimentally accessible  kinematic regime.
The $q^2$ dependence of the longitudinal cross section of charged pion electroproduction 
at JLab Hall C Ref.~\cite{Horn:2007ug} seems to be consistent with the predictions 
of the leading-twist collinear factorized description already at rather low values 
of the photon virtuality. However, the transverse cross section is large and its kinematic
dependence differs considerably from the scaling expectation. 
The more recent neutral pion electroproduction data from JLab Hall A and Hall B 
\cite{Collaboration:2010kna,Bedlinskiy:2014tvi} 
also suggest a large contribution of transversely polarized photons to the cross section,
different from the leading-twist formalism that predicts dominance of 
the longitudinal cross section.
Bringing evidence for the validity of QCD factorization
for $p \bar{p}$ annihilation into $e^+ e^- \pi^0$
in terms of $\pi N$ TDAs will suffer from the same
difficulties as the above-mentioned analysis of
hard electroproduction of pions at JLab.
 
Assuming the validity of the leading order factorized description 
for the signal reaction and adopting a particular normalization for $\pi N$ 
TDAs, we show the feasibility of measuring  of $p \bar{p} \to e^+ e^- \pi^0$ 
with the \={P}ANDA detector in the kinematic region where factorization
is expected to hold. With the cross section obtained from the simulations
we perform simple tests of pQCD at the leading twist, ignoring any higher-order effect.
This includes the measurement of the scaling laws by fitting the $q^2$ differential
cross sections and the determination of angular distributions by fitting the 
$\cos\theta^*_{\ell}$ cross sections.  

\begin{figure*}[hbtp]
\begin{center}
\epsfig{file=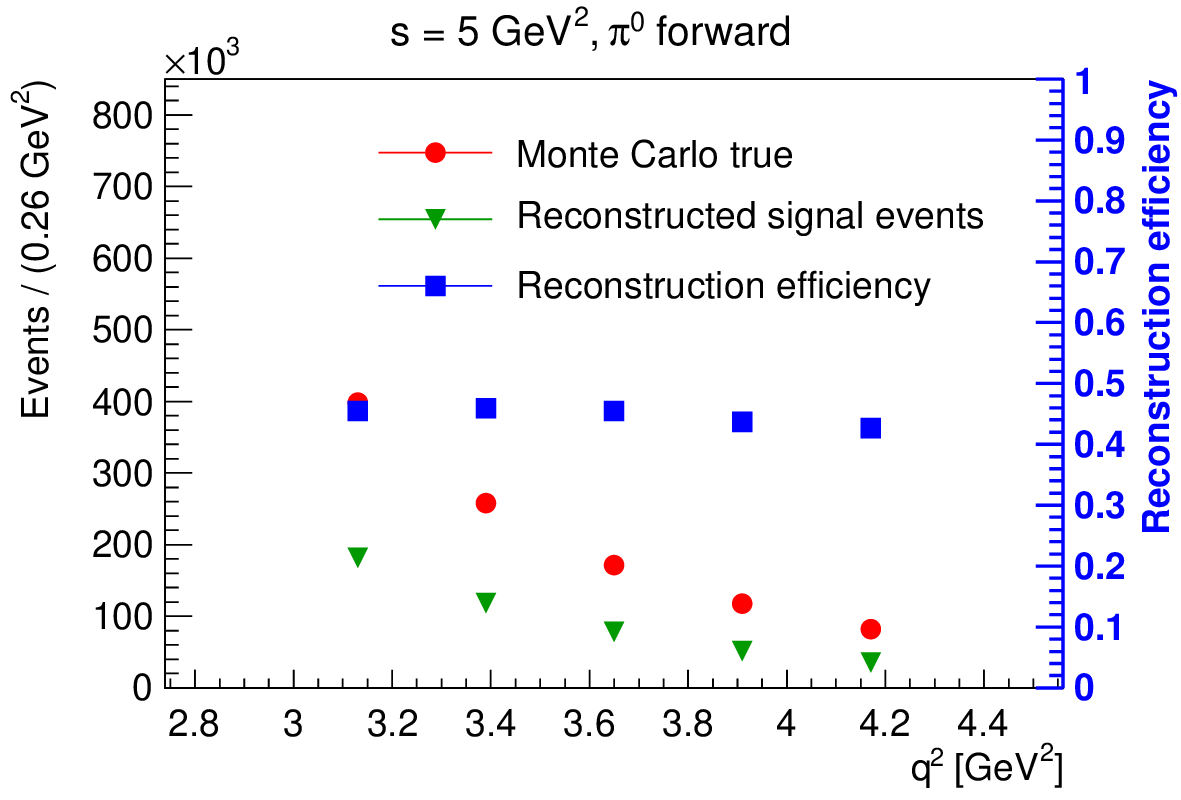,width=8.5cm}
\hspace*{0.5cm}
\epsfig{file=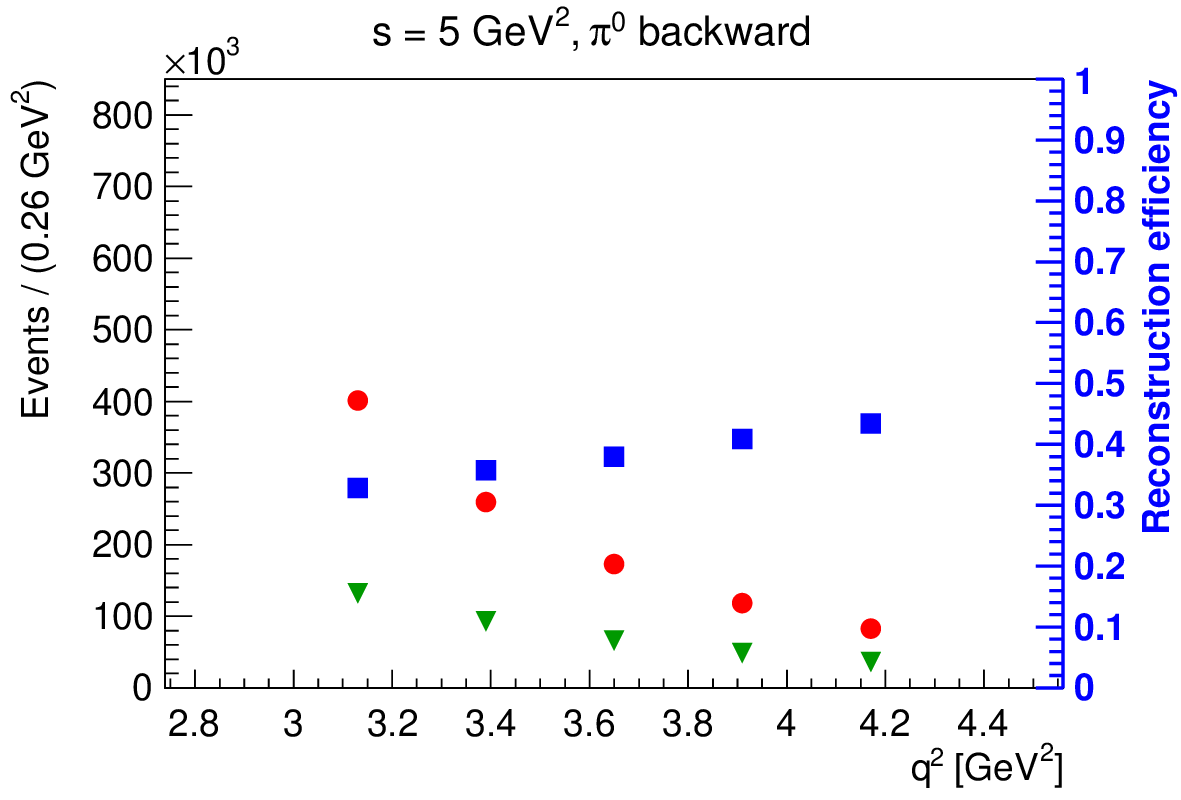,width=8.5cm} \\
\epsfig{file=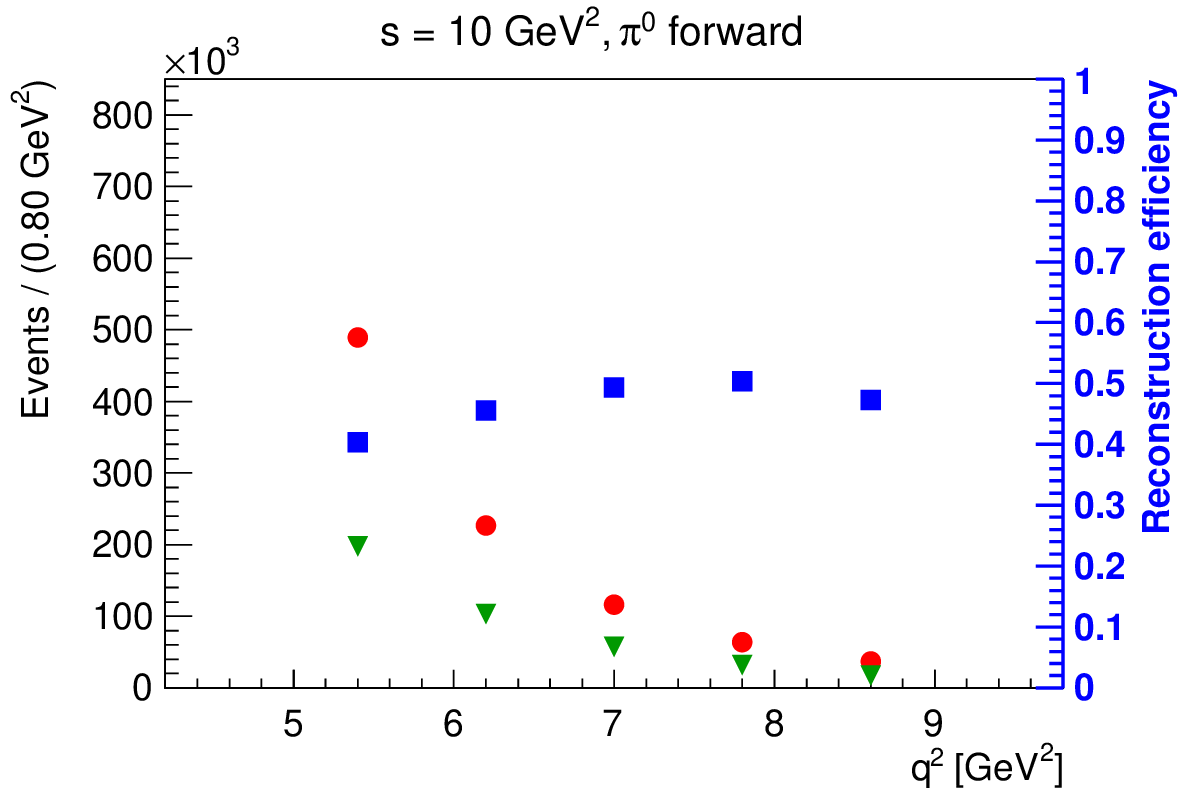,width=8.5cm}
\hspace*{0.5cm}
\epsfig{file=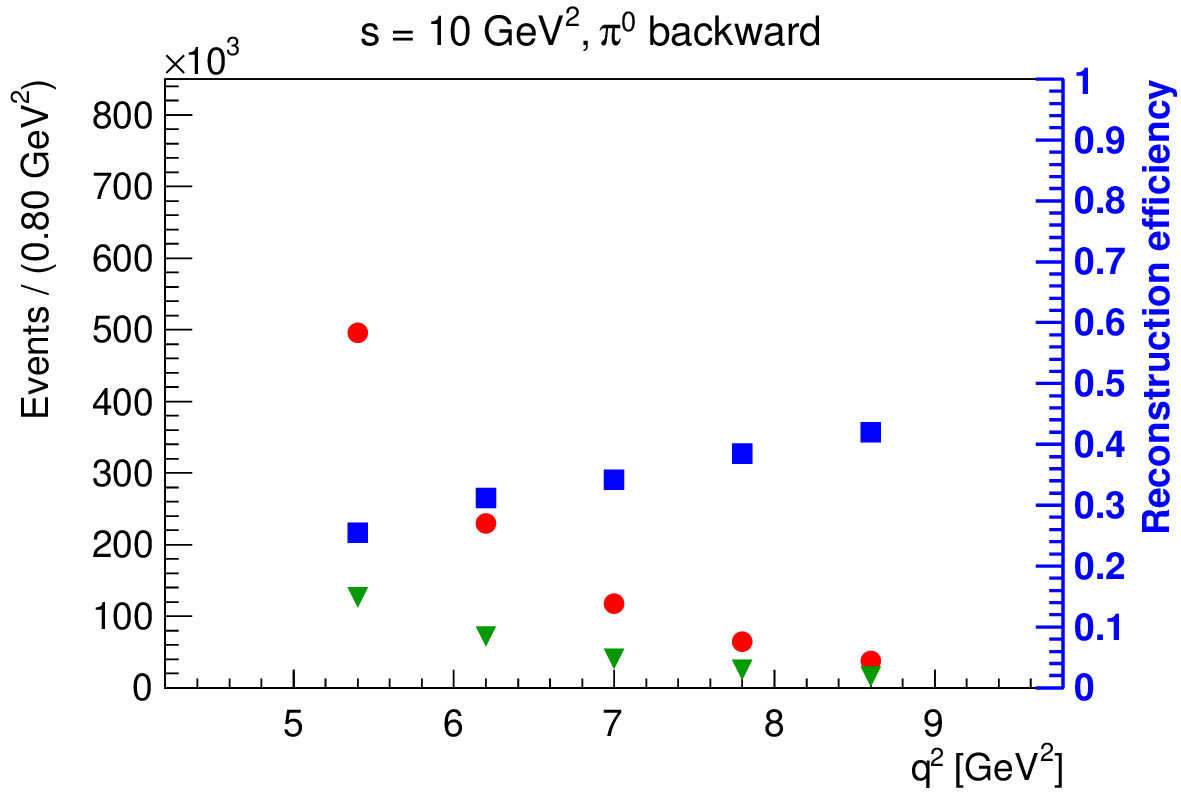,width=8.5cm}
\caption
[\protect{}]
{
Monte Carlo true $q^2$ distribution for signal events (red dots),
reconstructed signal events after event selection (green triangles)
and signal reconstruction efficiency (blue squares) as a function of $q^2$
for $s = 5~\rm GeV^{2}$ and $s = 10~\rm GeV^{2}$,
in both the $t$- ($\pi^0$ forward) and the $u$- ($\pi^0$ backward) channel kinematic regimes
determined using independent statistical samples of $10^6$ generated events.
}
\label{fig:efficiencies}
\end{center}
\end{figure*}

\section{Event selection}
\label{sec:event_selection}
Several simulations at the center of mass energy squared
$s = 5~\text{GeV}^2$ and $s = 10~\text{GeV}^2$ were done
using both simulated signal and background samples
in order to determine signal reconstruction efficiency,
background rejection power and feasibility of measuring
the differential cross section for $\bar{p}p\rightarrow e^+e^-\pi^0$
with an integrated luminosity of $2~\text{fb}^{-1}$.

The analysis procedure for the reconstruction of signal events
was designed by tuning the selection cuts in a way that
the signal to background ratio
was kept to its maximum value in the kinematic region
of the measurement. 
The selection strategy is mainly based on PID cuts.
In addition, kinematic fits were used to improve the measurement 
of the reconstructed momentum and energy of the particles.
The reconstruction of $e^+e^-\pi^0$ candidates
was done according to the following criteria:
\begin{itemize}
 \item the event contains exactly two charged tracks of opposite sign;
\item the particle associated to the negative track
 is identified by the PID software
 as an electron with minimum combined probability of $99\%$
 and, at least, with a minimum probability of $10\%$
 from each subdetector in \={P}ANDA;
\item the particle associated to the positive track
 is identified by the PID software
 as a positron with minimum combined probability of $99\%$
 and, at least, with a minimum probability of $10\%$
 from each subdetector in \={P}ANDA;
\item in the event, two photon candidates are reconstructed
from two energy deposits in the EMC
with a photon energy threshold $E_{\gamma} > 0.03$ GeV and no
track associated, and combined to give a $\pi^0$ candidate
with an invariant mass $0.115 < M(\gamma,\gamma) < 0.150$ GeV.
\end{itemize}

\noindent
At $s = 5~\text{GeV}^2$ and $s = 10~\text{GeV}^2$,
signal events were measured in the kinematic range
$3.0 < q^2 < 4.3~\text{GeV}^2$ and $5< q^2 < 9~\text{GeV}^2$, respectively.
In both cases, a $\pi^0$ candidate was reconstructed
in the forward or backward region $|\cos\theta_{\pi^0}| > 0.5$,
where the polar angle of the neutral pion is measured
with respect to the direction of the antiproton in
the $\bar{p}p$ CM system.
The kinematic region of the measurement ensures that,
at each $(q^2,\cos\theta_{\pi^0})$ point of the phase space,
the appropiate momentum transfer squared
($t$ or $u$ for the forward and backward pion production, respectively)
remains below $10\%$ of the $q^2$ value.
This is the definition adopted in this analysis of 
$|t| \ll q^2$ and $|u| \ll q^2$, needed to preserve the applicability of
the QCD collinear factorization description.

\section{Signal reconstruction efficiency}
\label{sec:efficiency}
\begin{figure*}[hbtp]
\begin{center}
\epsfig{file=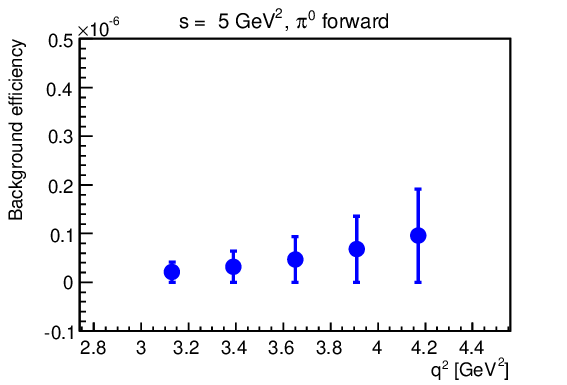,width=8.5cm}
\hspace*{0.5cm}
\epsfig{file=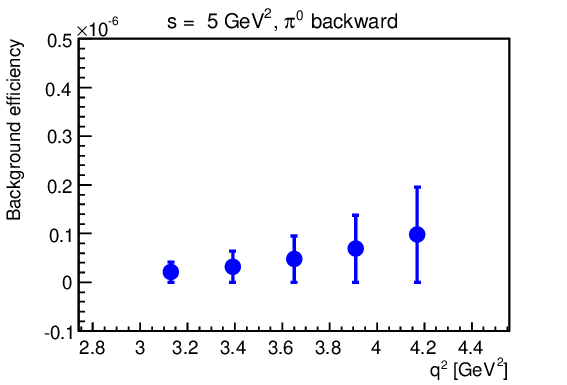,width=8.5cm} \\
\epsfig{file=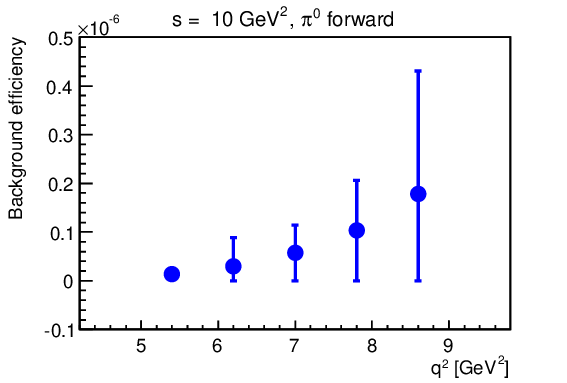,width=8.5cm}
\hspace*{0.5cm}
\epsfig{file=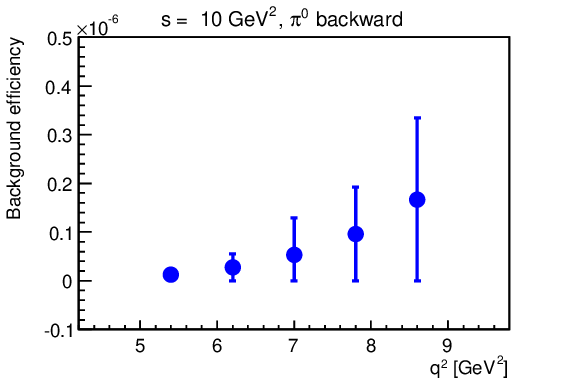,width=8.5cm}
\caption
[\protect{}]
{
Upper limit for the background reconstruction efficiency
at the confidence level of $67.3\%$ as a function of $q^2$
for $s = 5~\rm GeV^{2}$ and $s = 10~\rm GeV^{2}$,
in both the $t$- ($\pi^0$ forward) and the $u$- ($\pi^0$ backward) channel kinematic regimes
determined using independent statistical samples of $10^8$ generated events.
}
\label{fig:background_efficiencies}
\end{center}
\end{figure*}
High statistics simulations were done for the signal channel
$\bar{p}p\rightarrow e^+e^-\pi^0$
in order to determine the efficiency factors
needed to correct raw data for detector effects,
including efficiency in the reconstruction and bin migrations.
On the basis of a full Monte Carlo simulation,
the reconstruction efficiency measured in a given bin
of a generic observable $X$ is commonly defined as
$\epsilon = N^R/N^G$, where $N^R$ and $N^G$
are the number of reconstructed and generated events found in that bin,
with standard deviation
$\Delta\epsilon=\sqrt{N^R}/N^G$
assuming a Poisson distribution.
In order to determine the signal reconstruction efficiencies
as a function of $q^2$,
two full Monte Carlo simulations using $10^6$ generated events
each were performed at the center of mass energy squared
$s = 5~\text{GeV}^2$ in the $q^2$ range $3.0 < q^2 < 4.3~\text{GeV}^2$,
one in the $t$-channel regime, with the neutral pion in the forward region,
and another one in the $u$-channel regime, with the neutral pion in the backward region.
In an analogous way, two additional full simulations with the
same statistics were performed at $s = 10~\text{GeV}^2$,
in the range $5 < q^2 < 9~\text{GeV}^2$ also
for both the $t$- and the $u$- channel regimes.
The obtained reconstruction efficiencies in bins of $q^2$
are shown in Fig.~\ref{fig:efficiencies} for all four cases.
At $s = 5~\text{GeV}^2$ the reconstruction efficiency
shows a stable behaviour in $q^2$, with an almost constant value
around $45\%$ in the $t$-channel regime, whereas in the $u$-channel regime
the efficiency exhibits an increasing pattern from $33\%$ to $43\%$
in the $q^2$ range. At $s = 10~\text{GeV}^2$ a similar behaviour
is observed, with a mean value of $45\%$ in the $t$-channel regime
and increasing the reconstruction efficiency from $25\%$ to $40\%$ with $q^2$
in the $u$-channel regime.

\begin{figure*}[hbtp]
\begin{center}
\epsfig{file=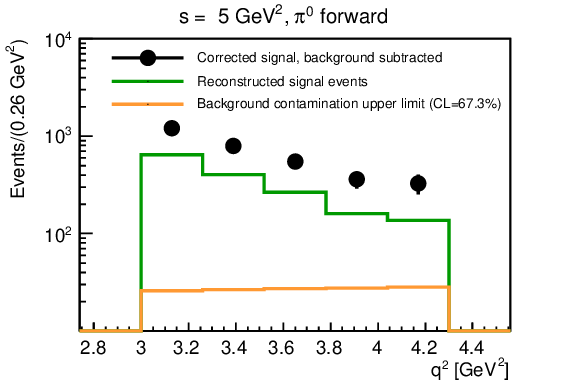,width=8.5cm}
\hspace*{0.5cm}
\epsfig{file=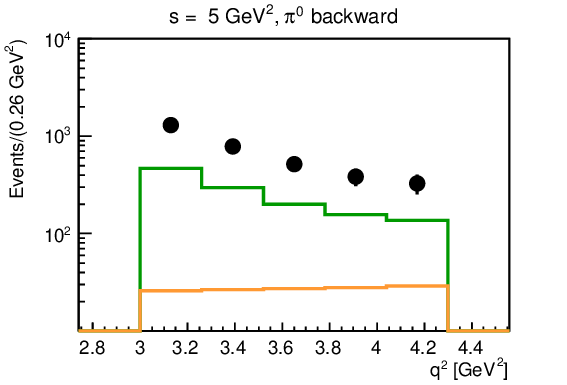,width=8.5cm} \\
\epsfig{file=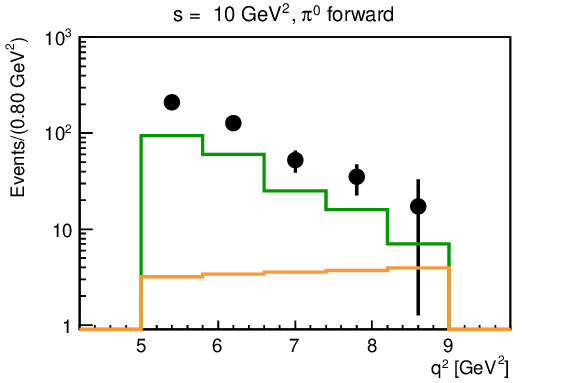,width=8.5cm}
\hspace*{0.5cm}
\epsfig{file=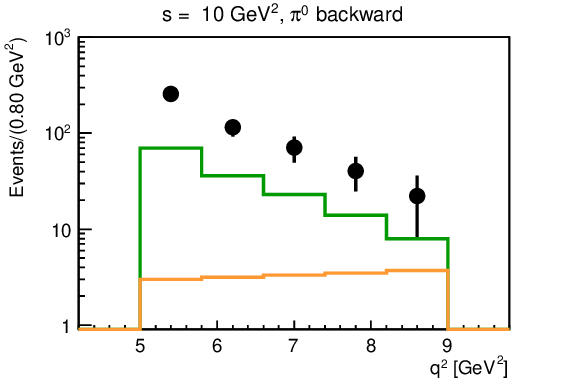,width=8.5cm}
\caption
[\protect{}]
{
The reconstructed signal after event selection (green),
the expected upper limit background contamination
at the $67.3\%$ of confidence level (orange)
and the reconstructed, efficiency-corrected signal after background subtraction
(black) in bins of $q^2$, for $s=5 ~\rm GeV^2$ and $s=10 ~\rm GeV^2$
in both the $t$- ($\pi^0$ forward) and the $u$- ($\pi^0$ backward) channel kinematic regimes
using statistical samples of integrated luminosity
$\mathcal{L} = 2 ~\rm fb^{-1}$.
}
\label{fig:corrected_and_subtracted}
\end{center}
\end{figure*}

\section{Background suppression}
\label{sec:background}
Analogous simulations to the ones described in
Section~\ref{sec:efficiency} were performed using samples of $10^8$
$\bar{p}p\rightarrow \pi^+\pi^-\pi^0$ generated
events in order to measure the background suppression power
achieved by the selection criteria defined
in Section~\ref{sec:event_selection}.
At $s=5$  $\text{GeV}^2$ and for both the $t$- and the $u$-channel regimes,
no pions were found after event selection.
At $s=10$ $\text{GeV}^2$, four $\pi^+\pi^-\pi^0$ events
were misidentified as $e^+e^-\pi^0$ events in the $t$-channel regime,
whereas in the $u$-channel regime only one background event survived the cuts.
The background suppression factor is defined as
the inverse of the probability that a $\pi^+\pi^-\pi^0$ event is misidentified
as a $e^+e^-\pi^0$ event. This probability can in fact be measured as the
``efficiency'' in the reconstruction of background events
when a $\pi^+\pi^-\pi^0$ sample is filtered by an algorithm
designed to reconstruct $e^+e^-\pi^0$ events.
For this reason, we denote this probability as $\epsilon_B$.
In situations of high suppression like this one,
where only a few (or even no event) are (is) reconstructed
in a given bin, the standard estimation of efficiency
and its error based on binomial or Poisson distributions
gives results in contradiction with intuition.
For instance, if no pion event is reconstructed in a given bin,
the value for the efficiency would be zero
with complete certainty (zero error) according to the Poisson distribution.
We can still in this case estimate
an \emph{upper limit} for $\epsilon_B$
at some value of confidence level,
depending on the available statistics.
In this analysis, to measure the reconstruction efficiency and its error,
a Bayesian approach which exhibits reasonable behaviour in the limit of high
suppression has been used to treat the background channel
(see Ref. \cite{Ullrich} for a review).
At the $67.3\%$ of confidence level (i.e. one sigma)
estimators of the upper limit of $\epsilon_B$
and its standard deviation $\Delta\epsilon_B$ are given
by the relations~\cite{Ullrich}:
\begin{equation}
\begin{split}
\label{eq:background_efficiency}
 \epsilon_B &= \frac{N^R + 1}{N^G + 2} \\
 \Delta\epsilon_B &=
 \sqrt{ \left(\frac{N^R + 1}{N^G + 2}\right)
   \left\{ \frac{N^R + 2}{N^G + 3} -\frac{N^R + 1}{N^G + 2}  \right\} }~.
\end{split}
\end{equation}
For the two energies simulated and in both, the $t$- and the $u$-channel regimes,
misidentification probabilities in bins of $q^2$
have been estimated in this way and are displayed
in Fig.~\ref{fig:background_efficiencies}.
The inverse $1/\epsilon_B$ then yields the suppression factor.
At $s=5~\text{GeV}^2$, the simulations show that
the background suppression factor goes from $5 \cdot 10^7$ at low $q^2$
down to $1 \cdot 10^7$ at large $q^2$.
At $s=10~\text{GeV}^2$,the background suppression factor goes
from $1 \cdot 10^8$ at low $q^2$ down to $6 \cdot 10^6$ at large $q^2$.
Under the assumption of a background to signal cross section ratio
$\sigma(\bar{p}p \rightarrow \pi^+\pi^-\pi^0) /
\sigma(\bar{p}p \rightarrow e^+e^-\pi^0) = 10^6$,
this means that the background pollution in a signal sample
will remain at the level of a few percent after event selection
for low $q^2$, whereas at larger values of $q^2$
it can be kept below $20\%$.
In case the cross section ratio is much larger than $10^6$,
a better background suppression can be achieved at the cost
of reducing the signal efficiency.
The estimated upper limit of background pollution
in a signal sample, necessary for the subsequent statistical subtraction,
is discussed in detail in Appendix \ref{appendix:background_subtraction}.

\section{Feasibility of measuring the $\mathsf{\bar{p}p\rightarrow e^+e^-\pi^0}$
differential cross section using an integrated luminosity $\mathsf{\mathcal{L}=2~\text{fb}^{-1}}$}
\label{sec:measurement}
\begin{table}
\begin{center}
\caption{
The differential cross section obtained from the simulations
$(d\sigma/dq^2)_{\rm sim}$ and its statistical error $\Delta_{\rm stat}$
in bins of $q^2$, compared to the input cross section in the Monte Carlo
$(d\sigma / dq^2)_{\rm MC}$, for $s=5 ~\rm GeV^2$  in both the $t$- and the
$u$-channel kinematic regimes. In each $q^2$ bin, the cross section is integrated 
in $|\cos\theta_{\pi^0}| > 0.5$.
}
\label{table:sigma_5}
\centerline{\normalsize $s = 5 ~\rm GeV^2$, $t$-channel ($\pi^0$ forward)}
\begin{tabular}{c c c c}
\hline\noalign{\smallskip}
$q^2$ bin  &
$(d\sigma / dq^2)_{\rm sim}$ &
$\Delta_{\rm stat}$ &
$(d\sigma / dq^2)_{\rm MC}$ \\
($\rm GeV^2$) & ($\rm fb / GeV^2$) & ($\rm fb / GeV^2$) & ($\rm fb / GeV^2$) \\
\noalign{\smallskip}\hline\noalign{\smallskip}
3.00, 3.26  & 2584   & 140    &  2388  \\
3.26, 3.52  & 1682   & 132    &  1600  \\
3.52, 3.78  & 1152   & 131    &  1105  \\
3.78, 4.04  &  754   & 136    &  782   \\
4.04, 4.30  &  680   & 145    &  567   \\
\noalign{\smallskip}\hline
& & & \\
\end{tabular}
\centerline{\normalsize $s = 5 ~\rm GeV^2$, $u$-channel ($\pi^0$ backward)}
\begin{tabular}{c c c c}
\hline\noalign{\smallskip}
$q^2$ bin  &
$(d\sigma / dq^2)_{\rm sim}$ &
$\Delta_{\rm stat}$ &
$(d\sigma / dq^2)_{\rm MC}$ \\
($\rm GeV^2$) & ($\rm fb / GeV^2$) & ($\rm fb / GeV^2$) & ($\rm fb / GeV^2$) \\
\noalign{\smallskip}\hline\noalign{\smallskip}
3.00, 3.26  & 2605  & 186  & 2388  \\
3.26, 3.52  & 1591  & 166  & 1600  \\
3.52, 3.78  & 1048  & 157  & 1105  \\
3.78, 4.04  &  782  & 150  & 782   \\
4.04, 4.30  &  667  & 147  & 567   \\
\noalign{\smallskip}\hline
\end{tabular}
\end{center}
\end{table}
\begin{table}
\begin{center}
\caption{
The differential cross section obtained from the simulations
$(d\sigma/dq^2)_{\rm sim}$ and its statistical error $\Delta_{\rm stat}$
in bins of $q^2$, compared to the input cross section in the Monte Carlo
$(d\sigma / dq^2)_{\rm MC}$, for $s=10 ~\rm GeV^2$  in both the $t$- and the
$u$-channel kinematic regimes. In each $q^2$ bin, the cross section is integrated 
in $|\cos\theta_{\pi^0}| > 0.5$.
}
\label{table:sigma_10}
\centerline{\normalsize $s = 10 ~\rm GeV^2$, $t$-channel ($\pi^0$ forward)}
\begin{tabular}{c c c c}
\hline\noalign{\smallskip}
$q^2$ bin  &
$(d\sigma / dq^2)_{\rm sim}$ &
$\Delta_{\rm stat}$ &
$(d\sigma / dq^2)_{\rm MC}$ \\
($\rm GeV^2$) & ($\rm fb / GeV^2$) & ($\rm fb / GeV^2$) & ($\rm fb / GeV^2$) \\
\noalign{\smallskip}\hline\noalign{\smallskip}
5.0, 5.8  & 137  & 15   & 144  \\
5.8, 6.6  &  83  & 14   & 72   \\
6.6, 7.4  &  34  &  9   & 39   \\
7.4, 8.2  &  23  &  8   & 23   \\
8.2, 9.0  &  11  & 10   & 14   \\
\noalign{\smallskip}\hline
& & & \\
\end{tabular}
\centerline{\normalsize $s = 10 ~\rm GeV^2$, $u$-channel ($\pi^0$ backward)}
\begin{tabular}{c c c c}
\hline\noalign{\smallskip}
$q^2$ bin  &
$(d\sigma / dq^2)_{\rm sim}$ &
$\Delta_{\rm stat}$ &
$(d\sigma / dq^2)_{\rm MC}$ \\
($\rm GeV^2$) & ($\rm fb / GeV^2$) & ($\rm fb / GeV^2$) & ($\rm fb / GeV^2$) \\
\noalign{\smallskip}\hline\noalign{\smallskip}
5.0, 5.8  & 162   & 21   & 144  \\
5.8, 6.6  &  73   & 14   & 72   \\
6.6, 7.4  &  45   & 14   & 39   \\
7.4, 8.2  &  26   & 10   & 23   \\
8.2, 9.0  &  14   &  9   & 14   \\
\noalign{\smallskip}\hline
\end{tabular}
\end{center}
\end{table}
\begin{figure*}[hbtp]
\begin{center}
\epsfig{file=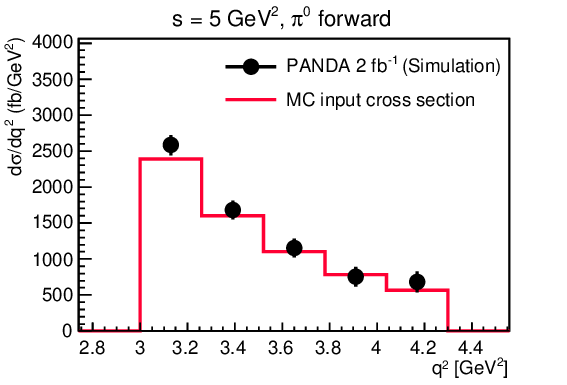,width=8.5cm}
\hspace*{0.5cm}
\epsfig{file=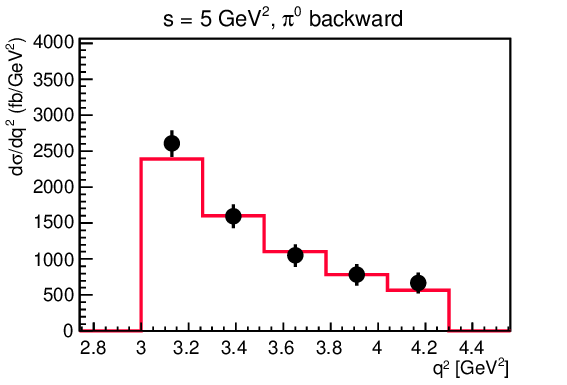,width=8.5cm} \\
\epsfig{file=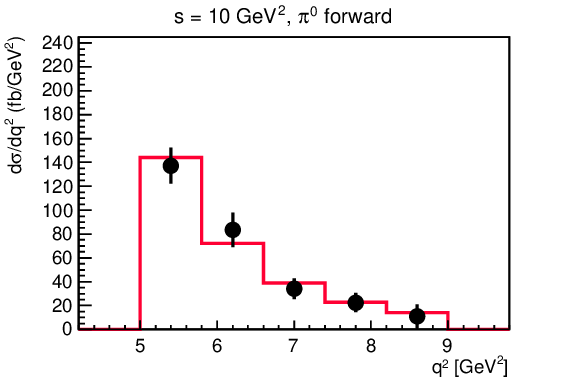,width=8.5cm}
\hspace*{0.5cm}
\epsfig{file=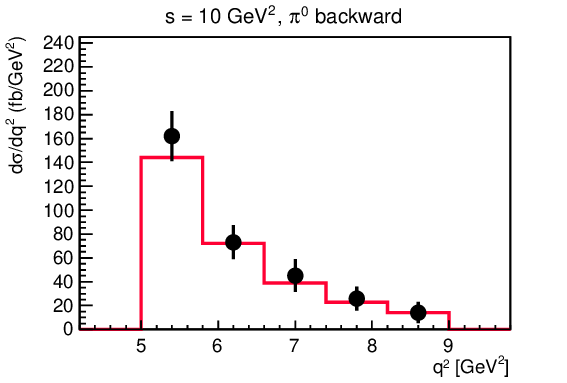,width=8.5cm}
\caption
[\protect{}]
{
The (background subtracted) $\bar{p}p \rightarrow e^+e^-\pi^0$
differential cross section from the simulation
$(d\sigma / dq^2)_{\rm sim}$ in bins of $q^2$
with a statistical sample of integrated luminosity
$\mathcal{L} = 2 ~\rm fb^{-1}$,
compared to the theoretical input in the Monte Carlo,
for $s=5 ~\rm GeV^2$ and $s=10 ~\rm GeV^2$,
in both the $t$- ($\pi^0$ forward) and the $u$- ($\pi^0$ backward) channel kinematic regimes.
}
\label{fig:sigma}
\end{center}
\end{figure*}

The feasibility of measuring the production
cross section for the signal channel $\bar{p}p\rightarrow e^+e^-\pi^0$
requires simulations using the expected statistics
corresponding to some particular value of integrated luminosity.
Running periods of six months with the average design luminosity of
$1.5 \cdot 10^{32} ~\text{cm}^{-2} ~\text{s}^{-1}$
will provide $2~\text{fb}^{-1}$
of integrated luminosity in \={P}ANDA~\cite{PANDAprogramm}.
In order to estimate the corresponding statistics,
we have first extrapolated the differential cross section
given by Eq.~(\ref{CS_working fromula}),
which corresponds to the limit of neutral pion with zero transverse momentum,
into the forward and backward cone $|\cos\theta_{\pi_0}|>0.5$.
Second, the extrapolated differential cross section
was integrated in the kinematic region of the measurement.
At $s=5$ $\text{GeV}^2$, integration in the range
$3.0 < q^2 < 4.3$ $\text{GeV}^2$ and $|\cos\theta_{\pi_0}|>0.5$
gave a value of $1675$ $\text{fb}$
for the integrated cross section.
At $s=10$ $\text{GeV}^2$, integration in the range
$5 < q^2 < 9$ $\text{GeV}^2$ and $|\cos\theta_{\pi_0}|>0.5$
gave a value of $233$ $\text{fb}$
for the integrated cross section.
Details on the extrapolation and integration of the differential
cross section in a two-dimensional bin
$(\Delta q^2, \Delta\cos\theta_{\pi_0})$
can be found in Appendix \ref{appendix:integration}.
The expected number of signal events in \={P}ANDA
using $\mathcal{L}=2~\text{fb}^{-1}$ are then $3350$ and $465$
at $s=5$ $\text{GeV}^2$ and $s=10$ $\text{GeV}^2$, respectively,
both in the $t$- and the $u$-channel kinematic regimes.
For each value of $s$, two full simulations have been performed
using these statistical samples in both channels.
Then, the raw reconstructed distributions have been
corrected bin by bin in $q^2$ with the efficiency factors $\epsilon$
determined by the high statistics
simulations described in Section~\ref{sec:efficiency}.
In addition, in each of the simulations and for each $q^2$ bin,
the remaining background contamination which would survive
the selection of signal events in a data sample of $2~\text{fb}^{-1}$
has been estimated. The estimation was done on the basis of
the background efficiency factors
discussed in Section~\ref{sec:background}
(Eq.~(\ref{eq:background_efficiency}))
and assuming a ratio
$\sigma(\bar{p}p\rightarrow \pi^+\pi^-\pi^0)/
\sigma(\bar{p}p\rightarrow e^+e^-\pi^0)=10^6$.
Consequently, the statistical error in the
number of reconstructed signal events $N^R$ has been corrected
to take into account the subtraction of the estimated
upper limit background contamination.
Details on the background subtraction procedure
are given in Appendix \ref{appendix:background_subtraction}.
An upper limit of background pollution at the level of a few percent
is expected at low $q^2$, remaining below $20\%$ at large values of $q^2$.
The raw reconstructed signal after event selection,
the expected upper limit background contamination,
and the efficiency-corrected signal after background subtraction
are shown in Fig.~\ref{fig:corrected_and_subtracted}
for the two energies simulated, in both the $t$- and the $u$-channel regimes.
The differential cross section obtained from the simulation
in a $q^2$ bin with width $\Delta q^2$
(integrated over $\cos\theta_{\pi^0}>0.5$ in the $t$-channel regime
and over $\cos\theta_{\pi^0} < - 0.5$ in the $u$-channel regime)
is then determined as:
\begin{equation}
 \left( \frac{d\sigma}{d q^2} \right)_{\text{sim}} =
 \frac{N^R}{\epsilon \cdot \mathcal{L} \cdot \Delta q^2} ~.
\end{equation}
The differential cross section obtained from the simulations $(d\sigma / dq^2)_{\text{sim}}$
in bins of $q^2$ together with the input cross section in the Monte Carlo
$(d\sigma / dq^2)_{\text{MC}}$ are shown in Tables~\ref{table:sigma_5}
and~\ref{table:sigma_10} and are displayed in Fig.~\ref{fig:sigma}.
For the comparison, the input cross section in the Monte Carlo,
which follows a $1/(q^2)^5$
distribution (see Eq.~(\ref{CS_working fromula})),
was normalized to the value of the integrated cross section
in the kinematic region of the measurement.
At $s=5$ $\text{GeV}^2$, the expected precision of the measurement
goes from $5\%$ at low $q^2$ to $21\%$ at high $q^2$
in the $t$-channel regime, 
and from $7\%$ to $22\%$ in the $u$-channel regime.
At $s=10$ $\text{GeV}^2$, the statistical error goes
from $11\%$ up to $91\%$ in the $t$-channel regime,
and from $13\%$ up to $64\%$ in the $u$-channel regime.
The results show that the signal channel identification
and background separation at $s=5$ $\text{GeV}^2$ is feasible,
with averaged statistical precision of $12\%$
(excluding the last $q^2$ bin with poor statistics).
At $s=10$ $\text{GeV}^2$, the lower statistics increases
the averaged uncertainty to $24\%$.

Bringing evidence for the consistency of the 
predictions by leading twist pQCD factorization with the experimentally measured
$p \bar{p}  \to \ell^+ \ell^- \pi^0$ cross section represents the
major goal of the proposed experimental studies. 
The harmonic analysis for separating the contribution of
the transversely polarized virtual photon as well as the 
$1/q^2$-scaling studies are the first crucial tests to be carried out to check 
the validity of the pQCD factorized description once sufficiently 
high quality experimental data appear.

\begin{figure*}[hbtp]
\begin{center}
\epsfig{file=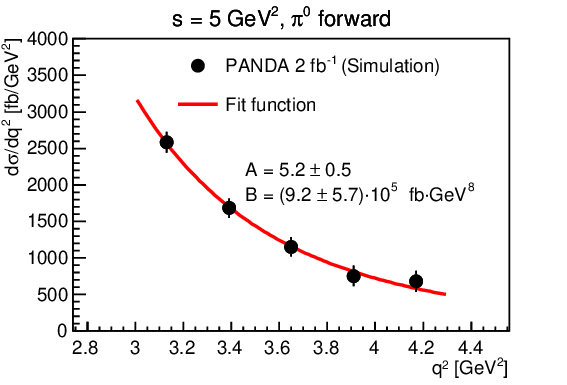,width=8.5cm}
\hspace*{0.5cm}
\epsfig{file=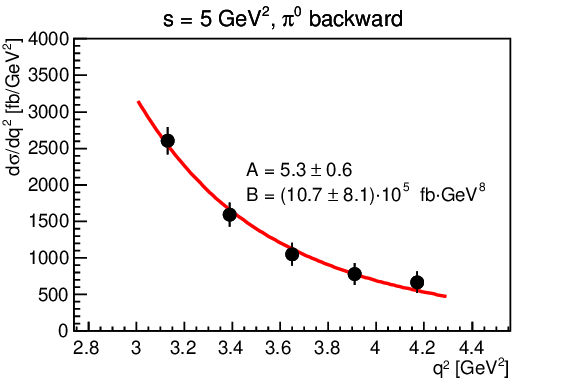,width=8.5cm} \\
\epsfig{file=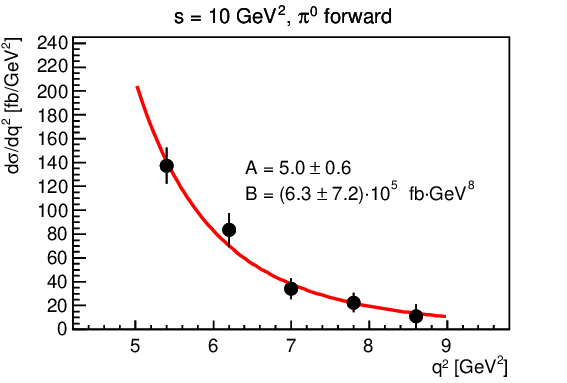,width=8.5cm}
\hspace*{0.5cm}
\epsfig{file=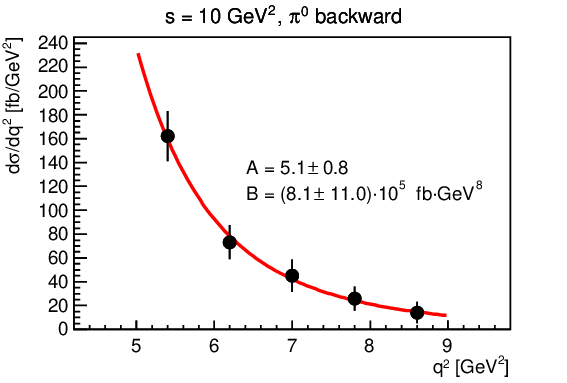,width=8.5cm} \\
\epsfig{file=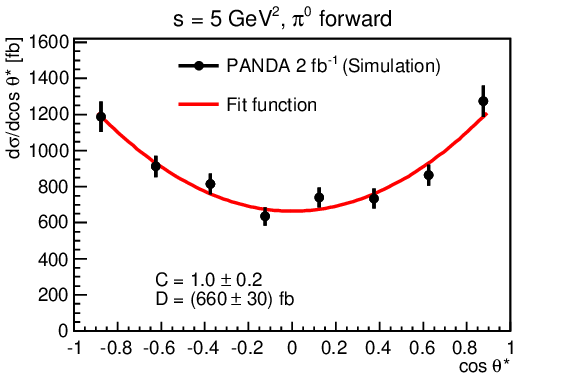,width=8.5cm}
\hspace*{0.5cm}
\epsfig{file=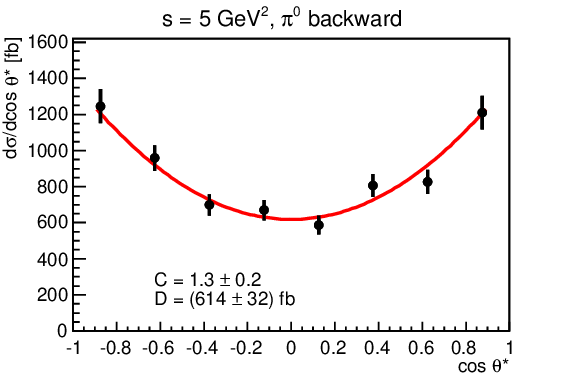,width=8.5cm} \\
\epsfig{file=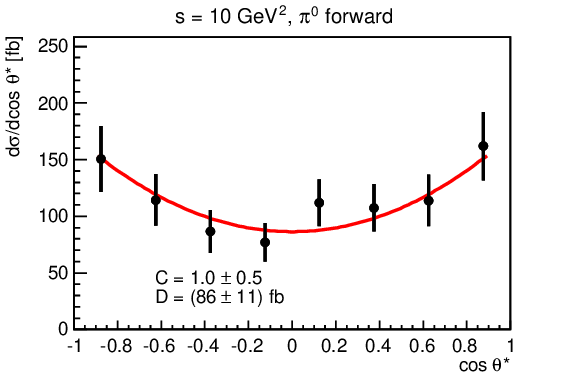,width=8.5cm}
\hspace*{0.5cm}
\epsfig{file=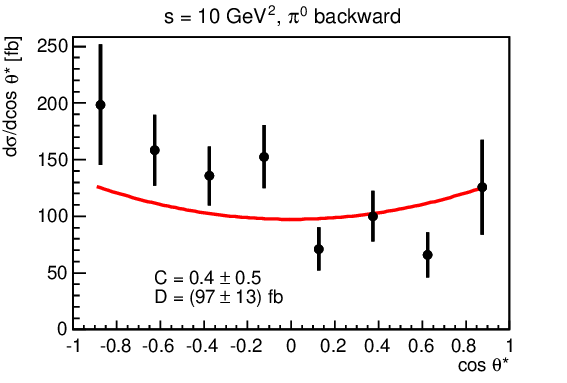,width=8.5cm}
\caption
[\protect{}]
{
The $\bar{p}p \rightarrow e^+e^-\pi^0$
differential cross sections obtained from the simulations (integrated luminosity
$\mathcal{L} = 2 ~\rm fb^{-1}$) for $s=5 ~\rm GeV^2$ and $s=10 ~\rm GeV^2$,
in both the $t$- ($\pi^0$ forward) and the $u$- ($\pi^0$ backward) 
channel kinematic regimes are fitted with the theoretical leading twist predictions.
For both the $q^2$ and $\cos\theta_{\ell}^*$ distributions, the fit function is
integrated over bin width. 
The average value for the  $q^2$ scaling exponent is $A = 5.2 \pm 0.3$.
The average value for the $\cos\theta_{\ell}^*$ prefactor is $C = 0.9 \pm 0.2$. 
}
\label{fig:fits}
\end{center}
\end{figure*}

In our Monte Carlo studies the scaling exponent is measured by fitting 
the $q^2$ distributions obtained from the simulations.
The fit function results from averaging the theoretical 
$d\sigma/dq^2$ in $q^2$ bins:
\begin{equation}
f(q^2) = \frac{1}{a}\int_{q^2 - a/2}^{q^2 + a/2} dx~ B ~\frac{1}{x^A} ~,
\end{equation}
which matches the measured observable. 
Here, the scaling parameter $A$ ($A=5.0$ is the input to the event generator)
and the normalization constant $B$ are the fit parameters
and $a$ is the $q^2$ bin width ($a = 0.26$ GeV$^2$ at $s = 5$ GeV$^2$ 
and $a = 0.80$  GeV$^2$ at $s = 10$ GeV$^2$).
The fitted $q^2$ distributions together with the measured values of 
$A$ and $B$ are displayed in Fig.~\ref{fig:fits} 
for the two values of energy simulated in both the $t$- and $u$-channel regimes.  
Using the measured values from all four fits, the scaling exponent $A$
has an average value of $5.2$ with a standard deviation of $0.3$.  
The large errors in the normalization constants obtained from the fits
are due to the strong correlation between the two fit parameters.

Repeating the same steps which led to the determination of 
the differential cross sections $d\sigma / dq^2$, 
as described at the beginning of this section,
the distributions $d\sigma / d\cos \theta_\ell^*$ 
have been also determined from the simulations.
The full kinematic range $-1 < \cos \theta_\ell^* < 1$
is covered with a total of $8$ bins for both energies and channels.  
At the leading twist and as a consequence 
of the dominance of the transverse polarization of the virtual photon, 
the cross section in $\cos \theta_\ell^*$ follows a distribution $(1 + \cos^2\theta_\ell^*)$.
The cross sections obtained from the simulations were then
fitted using the bin average of the theoretical expectation:
\begin{equation}
 g(\cos \theta_\ell^*) = \frac{1}{b}
 \int_{\cos\theta_\ell^* - b/2}^{\cos\theta_\ell^* + b/2} dx~ D (1 + C x^2)~.
\end{equation}
Here, the prefactor $C$ ($C=1$ is the input to the event generator) and 
the normalization constant $D$ are the fit parameters 
and $b = 0.25$ is the bin width.
The fitted $\cos \theta_\ell^*$ distributions together with the measured values
of $C$ and $D$ are shown in Fig.~\ref{fig:fits} for the two energies and channels.
Using the measured values from all four fits, the prefactor $C$
has an average value of $0.9$ with a standard deviation of $0.2$. 
The uncertainty in the prefactor $C$ contains the uncertainty 
in the reconstructed $q^2$, which is used to boost the $e^+$ and $e^-$ 
four momenta to the $\gamma^*$ rest frame in order to reconstruct 
the variable $\cos \theta_\ell^*$.   
Therefore, one expects to measure the $\cos \theta_\ell^*$ prefactor $C$
with less precision than the $q^2$ scaling exponent $A$.

\section{Conclusions and outlook}
\label{sec:conclusions}
In the framework of the \={P}ANDA@FAIR experiment,
cross section measurements of nucleon-antinucleon annihilation
into a highly virtual lepton pair in association with a pion
emitted in the forward or the backward region
will represent a novel test of the QCD collinear factorization approach
of hard exclusive reactions providing experimental access to the $\pi N$ TDAs.

In this paper we address the feasibility of measuring $\bar{p} p \to e^+ e^- \pi^0$
with the \={P}ANDA detector for the center of mass energy squared $s = 5$ GeV$^2$
and $s = 10$ GeV$^2$ for the kinematic regimes in which
the factorized description of the process
in terms of $\pi N$ TDAs and proton DAs can be assumed.
For $s = 5$ GeV$^2$, the kinematic region of the measurement was $3.0 < q^2 < 4.3$ GeV$^2$,
with the neutral pion scattered into the forward (or backward) cone selected by the condition
$| \cos \theta_{\pi^0}| > 0.5$. For $s = 10$ GeV$^2$, the kinematic region of the
measurement was $5 < q^2 < 9$ GeV$^2$, with $| \cos \theta_{\pi^0} | > 0.5$.

The input cross section for the event generator of signal events
is the leading twist, leading order calculation
which uses $\pi N$ TDAs and nucleon DAs
within the collinear factorization approach.
In our studies we employed the simple $\pi N$ TDA model
constraining $\pi N$ TDAs from chiral dynamics in terms of nucleon DAs.
This model is argued to provide a reliable normalization for $\pi N$
TDAs for the pion being produced exactly in the forward (backward) direction.
Therefore, this model  at least represents a reasonable first step approximation.
Future detailed feasibility studies will require the use
of a more sophisticated phenomenological model  proposed for $\pi N$ TDAs in 
\cite{Lansberg:2011aa} based on the spectral representation for baryon-to-meson TDAs 
in terms of quadruple distributions \cite{Pire:2010if}. 
Another possibility is given by the calculations of $\pi N$ TDAs
within the light-cone quark model approach \cite{Pasquini:2009ki}.

Our simulations at $s=5$ GeV$^2$ show that \={P}ANDA particle identification capabilities 
will allow a suppression of the hadronic background $\bar{p} p \to \pi^+ \pi^- \pi^0$
at the level of $5\cdot 10^7$ at low $q^2$, decreasing to $1\cdot 10^7$ for the larger values 
of $q^2$. At $s=10$ GeV$^2$, the suppression factor remains around $1\cdot 10^8$ at low $q^2$,
down to $6\cdot 10^6$ for large $q^2$.
For both energies, the signal reconstruction efficiency is kept at about $40\%$ 
on average. Consequently, we expect that the pion pollution in the signal sample
will remain at the level of a few percent at low $q^2$, and under control
below $20\%$ for larger values of four-momentum transfer squared.
The dedicated studies were performed with the statistics expected for an
integrated luminosity of $2$ fb$^{-1}$ and show that the future measurement
of the differential production cross section in bins of $q^2$ is feasible with
\={P}ANDA, with averaged statistical uncertainty of $12\%$  at $s = 5$ GeV$^2$,
and with averaged statistical uncertainty of $24\%$ at $s = 10$ GeV$^2$.
The cross sections obtained from the simulations in $q^2$ and $\cos\theta_{\ell}^*$
were also fitted to test pQCD factorization at the lowest order.
According to the simulations, the measured value for the  $q^2$ scaling exponent 
is $A = 5.2 \pm 0.3$. In the lepton angular distributions,
the measured value for the $\cos\theta_{\ell}^*$ prefactor is $C = 0.9 \pm 0.2$.
These results are promising concerning the experimental perspectives
for addressing the issue of validity of the pQCD factorized description 
of the $\bar{p} p \to \ell^+ \ell^- \pi^0$ reaction in terms of $\pi N$ TDAs and accessing 
$\pi N$ TDAs with \={P}ANDA. 
Other kinematic regions and other processes \cite{Pire:2013jva,GPS}
related to TDAs should be scrutinized in a similar way to evaluate their feasibility.
\begin{acknowledgement}
The success of this work relies critically on the expertise and dedication of the 
computing organizations that support \={P}ANDA.
We acknowledge  financial support from 
the Science and Technology Facilities Council (STFC), British funding agency, Great Britain;
the Bhabha Atomic Research Center (BARC) and the Indian Institute of Technology, Mumbai, India; 
the Bundesministerium f\"ur Bildung und Forschung (BMBF), Germany; 
the Carl-Zeiss-Stiftung 21-0563-2.8/122/1 and 21-0563-2.8/131/1, Mainz, Germany;
the Center for Advanced Radiation Technology (KVI-CART), Groningen, Netherlands;
the CNRS/IN2P3, France;
the Deutsche Forschungsgemeinschaft (DFG), Germany;
the Deutscher Akademischer Austauschdienst (DAAD), Germany;
the Forschungszentrum J\"ulich GmbH, J\"ulich, Germany;
the FP7 GA283286, European Commission funding;
the Ge\-sell\-schaft f\"ur Schwerionenforschung GmbH (GSI), Darmstadt, Germany;
the Helmholtz-Gemeinschaft Deutscher For\-schungs\-zen\-tren (HGF), Germany;
the INTAS, European Commission funding;
the Institute of High Energy Physics (IHEP) and the Chinese Academy of Sciences, Beijing, China;
the Istituto Nazionale di Fisica Nucleare (INFN), Italy;
the Ministerio de Educaci\'on y Ciencia (MEC) FPA2006-12120-C03-02, 
the Instituto de Fisica Corpuscular (IFIC) and the Universidad de Valencia-CSIC, Paterna, Spain;
the National Science Center of Poland (NCN)
and the Polish Ministry of Science and Higher Education (MNiSW), Poland;
the State Atomic Energy Corporation Rosatom and National Research Center ``Kurchatov Institute'', 
Russia;
the Schweizerischer Nationalfonds zur Forderung der wissenschaftlichen Forschung (SNF), Switzerland;
the Stefan Meyer Institut f\"ur Subatomare Physik and the 
\"Osterreichische Akademie der Wissenschaften, Wien, Austria;
the Swedish Research Council, Sweden.

We would like to thank B. Pire, L. Szymanowski and J.P. Lansberg for constructive 
and stimulating discussions, lots of explanations on the
topic of the TDAs and for a careful reading of the manuscript.
Finally, we are very grateful to D. Djukanovic for his invaluable
work in maintaining the dedicated HIMSTER cluster at
Helmholtz-Institut Mainz, where the simulations of this paper have been performed.
\end{acknowledgement}

\begin{appendices}
\section{Integration in a $\mathsf{(\Delta q^2, \Delta\cos\theta_{\pi^0})}$ bin}
\label{appendix:integration}
In this appendix we describe the extrapolation and integration
of the differential cross section (\ref{CS_working fromula}) in the
kinematic region of the measurement defined by the
two-di\-men\-sio\-nal bin $(\Delta q^2, \Delta\cos\theta_{\pi^0})$.

The kinematics of $\bar{p}(p_1) p(p_2) \rightarrow \gamma^*(q) \pi^0(k_3)$
is most easily solved in the CM frame, where the total three-momentum
of both the initial and final state is zero.
By convention, the direction of the antiproton defines
the positive $z$ direction of the coordinate system.
Also by convention, the $x$ axis of the coordinate system
is chosen to be perpendicular to the scattering plane.
With this choice, the four-momenta of the initial-
and final-state particles become
\begin{equation}
\label{eq:four_momenta}
\begin{split}
 p_1 &= (E, 0, 0, k_i)  \\
 p_2 &= (E, 0, 0, -k_i) \\
 q &= (E_{\gamma}, 0, -k_f\sin\theta_{\pi^0}, -k_f\cos\theta_{\pi^0}) \\
 k_3 &= (E_{\pi^0}, 0, k_f\sin\theta_{\pi^0}, k_f\cos\theta_{\pi^0}) ~,
\end{split}
\end{equation}
with energies given by $E = \sqrt{M^2 + k_i^2}$,
$E_{\gamma} = \sqrt{q^2 + k_f^2}$ and $E_{\pi^0} = \sqrt{m_{\pi^0}^2 + k_f^2}$.
The condition $2E = \sqrt{s}$ fixes the three-momentum modulus squared
of both proton and antiproton to be
\begin{equation}
 k_i^2 = \frac{1}{4} (s-4M^2)~.
\end{equation}
In the same way, the energy conservation relation in the final state
$E_{\gamma} + E_{\pi^0} = \sqrt{s}$ fixes the momenta of virtual photon and
neutral pion, with the result
\begin{equation}
 k_f^2 = \frac{1}{4s}
 \big[ s^2 -2(q^2 + m_{\pi^0}^2)s + (q^2 - m_{\pi^0}^2)^2 \big]~.
\end{equation}
Using the four-momenta given by Eq. (\ref{eq:four_momenta}),
the antiproton to pion four-momentum transfer squared $t\equiv (p_1 - k_3)^2$
is given by
\begin{equation}
\begin{split}
 t &= (p_1 - k_3) ^2 \\
   &= p_1^2 +k_3^2 -2 p_1 \cdot k_3 \\
   &= M^2 + m_{\pi^0}^2  -2 ( E E_{\pi^0} - k_i k_f \cos\theta_{\pi^0} ) ~,
\end{split}
\end{equation}
which can be written as
\begin{equation}
\begin{split}
\label{eq:t}
 t = \frac{1}{2} \big[
   m_{\pi^0}^2 &+\cos\theta_{\pi^0} \sqrt{1-4M^2/s} ~~\Lambda(s,q^2,m_{\pi^0}^2) \\
   &+2M^2 +q^2 -s \big] ~,
\end{split}
\end{equation}
with $\Lambda(x,y,z) \equiv \sqrt{x^2 + y^2 +z^2 -2xy -2xz -2yz}$.
Eq. (\ref{eq:t}) expresses the dependence of the variable $t$
on $q^2$ and $\cos\theta_{\pi^0}$ for a given value of the center of mass energy squared $s$.

The integration of the leptonic phase space degrees of freedom
in Eq. (\ref{CS_working fromula}) is straightforward:
\begin{equation}
\begin{split}
\label{eq:sigma_int_lepton}
 \frac{d\sigma}{dt~dq^2} ~\bigg|_{\Delta_T=0} &=
 \int_{-1}^{1} d\cos\theta^*_\ell
 \frac{d\sigma}{dt~dq^2~d\cos\theta^*_\ell} ~\bigg|_{\Delta_T=0} \\
 &=\frac{8}{3} ~K ~\frac{1}{s -4M^2} ~\frac{1}{(q^2)^5}~.
\end{split}
\end{equation}
The integration of this equation in the two-dimensional bin
$(\Delta q^2,\Delta \cos\theta_{\pi^0})$
defined by the limits $q^2_1 < q^2 < q^2_2$ and
$\cos\theta_1 < \cos\theta_{\pi^0} < 1$ is done by first
mapping the $\cos\theta$ boundaries to the ($q^2$ dependent)
$t$ boundaries as given by Eq. (\ref{eq:t}):
$\cos\theta_1 < \cos\theta < 1 \Rightarrow t_{cut}(q^2) < t < t_{max}(q^2)$,
with $t_{cut}(q^2) \equiv t(\cos\theta_1,q^2)$
and $t_{max}(q^2) \equiv t(1,q^2)$.
Extrapolating the differential cross section (\ref{eq:sigma_int_lepton})
(obtained at $\cos\theta_{\pi^0}=1$) to the angular region $\cos\theta_1 < \cos\theta_{\pi^0} < 1$,
the integration in the two-dimensional bin $(\Delta q^2,\Delta \cos\theta_{\pi^0})$
is done as
\begin{equation}
\begin{split}
\label{eq:sigma_integrated}
 \sigma &= \int_{q^2_1}^{q^2_2} dq^2 \int_{t_{\rm cut}(q^2)}^{t_{\rm max}(q^2)} dt
 ~\frac{d\sigma}{dq^2 ~dt} ~\bigg|_{\Delta_T=0} \\
 &= \frac{8}{3} ~K ~\frac{1}{s-4M^2}
      ~\int_{q^2_1}^{q^2_2} dq^2 ~\frac{1}{(q^2)^5}
      \big[ t_{\rm max}(q^2) - t_{\rm cut}(q^2) \big] ~.
\end{split}
\end{equation}
For the estimates used in this analysis,
the $q^2$ integration in Eq. (\ref{eq:sigma_integrated})
has been done numerically.

\section{Background subtraction}
\label{appendix:background_subtraction}
In this appendix we describe the background subtraction procedure
carried out in our analysis.
For the clarity of the notation, we will refer to the signal channel
$\bar{p}p\rightarrow e^+e^-\pi^0$ with the subscript $S$,
and we will refer to the background channel
$\bar{p}p\rightarrow \pi^+\pi^-\pi^0$ with the subscript $B$.
In a data sample with luminosity $\mathcal{L}$,
the expected number of signal and background events
produced in an experiment is $N_S = \sigma_S~\mathcal{L}$
and $N_B = \sigma_B~\mathcal{L}$.
After the event selection procedure, the expected number of reconstructed
signal and background events becomes
$N_S^R = \epsilon_S~\sigma_S~\mathcal{L}$ and
$N_B^R = \epsilon_B~\sigma_B~\mathcal{L}$,
where $\epsilon_S$ and $\epsilon_B$ are the reconstruction
efficiencies for the signal and background channels.
We remark that due to the high background suppression,
$\epsilon_B$ is understood as an upper limit for the background
reconstruction efficiency estimated at some confidence level,
as discussed in Section~\ref{sec:background}.
Consequently, $N_B^R$ is also understood
as an upper limit for background contamination,
estimated at the same confidence level.
The total number of observed events in the sample used for the measurement
is therefore $N^R = N_S^R + N_B^R$, for which we assume
a Poisson distribution with standard deviation $\Delta N^R =\sqrt{N^R}$.
The estimation of the number of signal events in this sample
is then done by subtracting the remaining background contamination,
and assuming standard error propagation:
\begin{equation}
N_S^R = N^R - N_B^R, \quad
(\Delta N_S^R)^2 = (\Delta N^R)^2 + (\Delta N_B^R)^2 ~.
\end{equation}
The estimation of the background contamination
$N_B^R = \epsilon_B~\sigma_B~\mathcal{L}$
requires the knowledge of the cross section $\sigma_B$,
which can be measured with \={P}ANDA. In our analysis, however,
we simply assume the relation $\sigma_B = 10^6 ~\sigma_S$.
The computation of the standard deviation is also determined
by simple error propagation, according to
\begin{equation}
\left( \frac{\Delta N_B^R }{N_B^R} \right)^2 =
\left( \frac{\Delta \epsilon_B}{\epsilon_B} \right)^2 +
\left( \frac{\Delta \sigma_B}{\sigma_B} \right)^2 ~.
\end{equation}
Relying of the fact that the cross section corresponding to
three-pion production from $\bar{p}p$ annihilation will be measured at \={P}ANDA
with great precision due to its high statistics,
we have neglected the last term $\Delta \sigma_B / \sigma_B$ in our analysis.

\end{appendices}


\end{document}